\documentclass[%
 reprint,
 superscriptaddress,
 amsmath,amssymb,
 aps,
prb
]{revtex4-2}

\usepackage{graphicx}
\usepackage{dcolumn}
\usepackage{bm}
\usepackage{color}
\usepackage{xcolor}

\usepackage{amsfonts}
\usepackage{siunitx}
\usepackage{tabularx}

\newcommand{\curlyI}{\mathcal{I}}
\newcommand{\IH}{\curlyI_{\text{H}}}
\newcommand{\curlyL}{\mathcal{L}}
\newcommand{\curlyLTilde}{\tilde{\mathcal{L}}}

\newcommand{\curlyP}{\mathcal{P}}
\newcommand{\EH}{\curlyP_{\text{H}}}
\newcommand{\PH}{P_{\text{H}}}
\newcommand{\curlyC}{\mathcal{C}}
\newcommand{\curlyV}{\mathcal{V}}
\newcommand{\Gtwo}{ G^{(2)} }
\newcommand{\gtwo}{ g^{(2)}(0) }
\newcommand{\Gone}{ G^{(1)} }
\newcommand{\Gtwopop}{ G_\text{pop}^{(2)} }

\newcommand{\Gtwoi}{ G^{(2)}_{i} }
\newcommand{\Gonei}{ G^{(1)}_{i} }
\newcommand{\Gtwopopi}{ G_{\text{pop},i}^{(2)} }
\newcommand{\GtwoHOMi}{ G_{\text{HOM},i}^{(2)} }
\newcommand{\twophotmat}{\rho^{2 \gamma} }

\newcommand{\Efsp}{E_\text{fss}}
\newcommand{\Ebind}{E_\text{bind}}
\newcommand{\EX}{E_\text{X}}

\newcommand{\omegaC}{\omega_{\text{cav-centr}}}
\newcommand{\omegaSpl}{\omega_{\text{cav-split}}}
\newcommand{\omegaH}{\omega_{\text{H}}}
\newcommand{\omegaV}{\omega_{\text{V}}}
\newcommand{\omegaCav}{\omega_\text{cav}}
\newcommand{\omegaS}{\omega_{\text{S}}}
\newcommand{\epsiS}{\epsilon_{\text{S}}}
\newcommand{\muS}{\mu_{\text{S}}}

\newcommand{\DeltaXHG}{\Delta_{\text{Cav,X}_{\text{H}}\to\text{G}}}

\newcommand{\GammaBXH}{\Gamma^{\text{Cav,}\SI{4.2}{\K}}_{\text{XX}\to\text{X}_{\text{H}}}}
\newcommand{\GammaXHG}{\Gamma^{\text{Cav,}\SI{4.2}{\K}}_{\text{X}_{\text{H}}\to\text{G}}}

\newcommand{\D}{\mathrm{d}}
\newcommand{\E}{\mathrm{e}}
\newcommand{\I}{\mathrm{i}}
\newcommand{\hc}{\text{h.c.}}
\newcommand{\AvB}{\langle B \rangle}
\newcommand{\ketbra}[2]{|#1\rangle\negthinspace\langle#2|}
\newcommand{\ket}[1]{|#1\rangle}

\newcommand{\gX}{g^{\rm X}_\omega}
\newcommand{\gB}{g^{\rm XX}_\omega}
\newcommand{\bX}{b^{\rm X}_\omega}
\newcommand{\bB}{b^{\rm XX}_\omega}
\newcommand{\gsigma}{g^{\hat{\sigma}}}
\newcommand{\gsigmaConti}{g^{\hat{\sigma}}_\omega}
\newcommand{\Gsigma}{G^{\hat{\sigma}}_\omega}
\newcommand{\GsigmaTilde}{\tilde{G}^{\hat{\sigma}}_\omega}

\newcommand{\asigma}{a^{\hat{\sigma}}}
\newcommand{\acav}{a_{\omegaCav}}

\newcommand{\bsigma}{b^{\hat{\sigma}}_\omega}
\newcommand{\kappasigma}{\kappa^{\hat{\sigma}}_\omega}
\newcommand{\kappacav}{\kappa_{\omegaCav}}

\newcommand{\Kappasigma}{K^{\hat{\sigma}}_\omega}
\newcommand{\omegasigma}{\omega^{\hat{\sigma}}}
\newcommand{\omegaQDT}{\omega_{\text{QD-trans}}}

\newcommand{\TabRef}[1]{Table~\ref{#1}}
\newcommand{\Fig}[1]{Fig.~#1}
\newcommand{\Figref}[1]{Fig.~\ref{#1}}

\newcommand{\FIGref}[1]{Figure~\ref{#1}}

\newcommand{\Eq}[1]{Eq.~#1}
\newcommand{\Eqref}[1]{Eq.~\eqref{#1}} 
\newcommand{\Eqsref}[1]{Eqs.~\eqref{#1}}

\newcommand{\Secref}[1]{Section~\ref{#1}}

\newcommand{\Appref}[1]{Appendix~\ref{#1}}
\newcommand{\Apprefs}[2]{Appendices~\ref{#1} and \ref{#2}}

\usepackage{tabularx}
\usepackage{enumerate}
\usepackage{enumitem}
\usepackage{verbatim}
\usepackage{hyperref}
\usepackage[nolist]{acronym}

\begin{document}

\preprint{APS/123-QED}

\title{High-quality single photons from cavity-enhanced biexciton-to-exciton transition} 

\author{Nils Heinisch}
 \affiliation{Department of Physics and Center for Optoelectronics and Photonics Paderborn (CeOPP), Paderborn University, Warburger Strasse 100, 33098 Paderborn, Germany}
 \affiliation{Institute for Photonic Quantum Systems (PhoQS), Paderborn University, 33098 Paderborn, Germany}

\author{Francesco Salusti}
\affiliation{Department of Physics and Center for Optoelectronics and Photonics Paderborn (CeOPP), Paderborn University, Warburger Strasse 100, 33098 Paderborn, Germany}%
\affiliation{Institute for Photonic Quantum Systems (PhoQS), Paderborn University, 33098 Paderborn, Germany}

\author{Mark R. Hogg}
\affiliation{Department of Physics, University of Basel, Klingelbergstrasse 82, 4056 Basel, Switzerland}%

\author{Timon L. Baltisberger}
\affiliation{Department of Physics, University of Basel, Klingelbergstrasse 82, 4056 Basel, Switzerland}%

\author{Malwina A. Marczak}
\affiliation{Department of Physics, University of Basel, Klingelbergstrasse 82, 4056 Basel, Switzerland}%

\author{Sascha R. Valentin}
\affiliation{Faculty of Physics and Astronomy, Experimental Physics VI, Ruhr University Bochum, Universitätsstrasse 150, 44801 Bochum, Germany}

\author{Arne Ludwig}
\affiliation{Faculty of Physics and Astronomy, Experimental Physics VI, Ruhr University Bochum, Universitätsstrasse 150, 44801 Bochum, Germany}

\author{Klaus D. J\"ons}
\affiliation{Department of Physics and Center for Optoelectronics and Photonics Paderborn (CeOPP), Paderborn University, Warburger Strasse 100, 33098 Paderborn, Germany}%
\affiliation{Institute for Photonic Quantum Systems (PhoQS), Paderborn University, 33098 Paderborn, Germany}

\author{Richard J. Warburton}
\affiliation{Department of Physics, University of Basel, Klingelbergstrasse 82, 4056 Basel, Switzerland}%

\author{Stefan Schumacher}
\affiliation{Department of Physics and Center for Optoelectronics and Photonics Paderborn (CeOPP), Paderborn University, Warburger Strasse 100, 33098 Paderborn, Germany}%
\affiliation{Institute for Photonic Quantum Systems (PhoQS), Paderborn University, 33098 Paderborn, Germany}
\affiliation{Wyant College of Optical Sciences, University of Arizona, Tucson, AZ 85721, USA}%

\date{\today}

\begin{abstract}
Resonant laser excitation of a two-level system with subsequent single-photon emission can be used to generate single photons with high indistinguishability or Hong-Ou-Mandel (HOM) visibility. However, spectral overlap between excitation laser and emitted photons generally poses significant challenges. Furthermore, emitter re-excitation intrinsically limits achievable single-photon purity. Established solutions mitigate these issues at significant cost to source efficiency and with increased source complexity. This motivates the use of few-level systems with spectral separation of excitation and emission pathways. One option is a three-level cascade. However, without targeted lifetime engineering of emitting states, the cascade naturally limits achievable photon indistinguishability. Here we study a semiconductor quantum dot with resonant and selective cavity-enhancement of biexciton-to-exciton transition. Following resonant two-photon excitation of the biexciton state, we collect the emitted single photon with the cavity. This approach circumvents emitter re-excitation and naturally introduces spectral separation of excitation laser and emitted single photon. Supported by first experimental results, we demonstrate theoretically that with selective Purcell enhancement, the observed quality quantifiers of single-photon emission (purity, equivalently $\gtwo$, and HOM visibility $\curlyV$, equivalently indistinguishability) are competitive with respect to high-quality deterministic quantum-dot single-photon sources. This is already achieved without systematic optimization or targeted system engineering, which firmly places the reported approach as a viable route to the next generation of highest-quality quantum-dot based deterministic single-photon sources.
\end{abstract}

\maketitle



\section{Introduction}

Semiconductor quantum dots (QDs) have garnered significant interest as deterministic sources of single photons, showing great promise for quantum information processing \cite{Somaschi_2016,Wang_2016,Uppu_2020_SciAdv,Tomm_2021,Ding_2025,Liu_2018}. Despite substantial progress, the efficient generation and extraction of high-quality photons -- crucial for practical applications -- remains a challenge \cite{Praschan_2022,IlesSmith_2017,Ramsay_2010,Huber_2013,Somaschi_2016,Wei_2014,Varnava_2008,Jennewein_2011}. Various excitation and extraction strategies have been proposed and explored \cite{Reindl_2019,Thomas_2021,Ardelt_2014,Stufler_2006,Bauch_2021,Bauch_2024,He_2019,Koong_2021,Vannucci_2023,SUPER,SUPERinAction,Boos_2024,Undeutsch_2025,Hanschke_2025,Vannucci_2025,Cardoso_2025,GonzalezRuiz_2025}, yet each introduces its own set of limitations \cite{IlesSmith_2017,Ramsay_2010,Huber_2013,Jonas_2022,Mueller_2015,Hanschke_2018,Fischer_2017,Fischer_2_2017}. 

Two major difficulties remain that a practical single‑photon source must overcome. Firstly, the source must generate highly indistinguishable photons (i.e., photons with high Hong-Ou-Mandel visibility) and simultaneously must emit precisely one photon per excitation cycle (i.e., unity single-photon purity). In the simplest excitation-emission scheme based on resonant excitation of a two-level emitter, single-photon purity is diminished by emitter re-excitation, especially for resonant excitation in combination with cavity enhancement of the same transition \cite{Hanschke_2018,Mueller_2015,Fischer_2017,Fischer_2_2017,Heinisch_2024}. Secondly, the highly indistinguishable single photons must be separated from the excitation pulse. However, many schemes use the same optical mode for excitation and single-photon emission. As a result, either the quality of the single photons is reduced, or single-photon extraction is only possible at the expense of overall reduced photon output \cite{GonzalezRuiz_2025}.

One widely used approach to overcome these limitations in semiconductor quantum dot systems involves initiating the photon generation from a biexciton (XX) state rather than from an exciton (X) state, thereby exploiting a cascaded transition for single‑photon emission \cite{Schoell_2020}. This configuration enables a clear separation -- both spectrally and in polarization -- between the generated photons and the laser while avoiding re‑excitation \cite{Mueller_2014}, even for straightforward excitation techniques such as resonant two-photon excitation (TPE) \cite{Schweickert_2018}. However, challenges remain, particularly concerning the XX-X lifetime ratio, which reduces the achievable Hong-Ou-Mandel visibility, and equivalently photon indistinguishability \cite{Schoell_2020,Simon_2005}. Common strategies to mitigate this issue generate single photons via the exciton‑ground (X-G) state transition, enhancing the XX-X transition using external stimulants such as lasers or cavities \cite{Hagen_2025,Karli_2024,Bauch_2024,Huber_2013,Behrends_2026}. However, photons emitted from the exciton may suffer from this intermediate step (e.g., by an increased time frame over which detrimental influences such as dephasing can act) \cite{Baltisberger_2025,Behrends_2026}.\\

\begin{figure*}[t]
\centering
\includegraphics[width=1.0\textwidth]{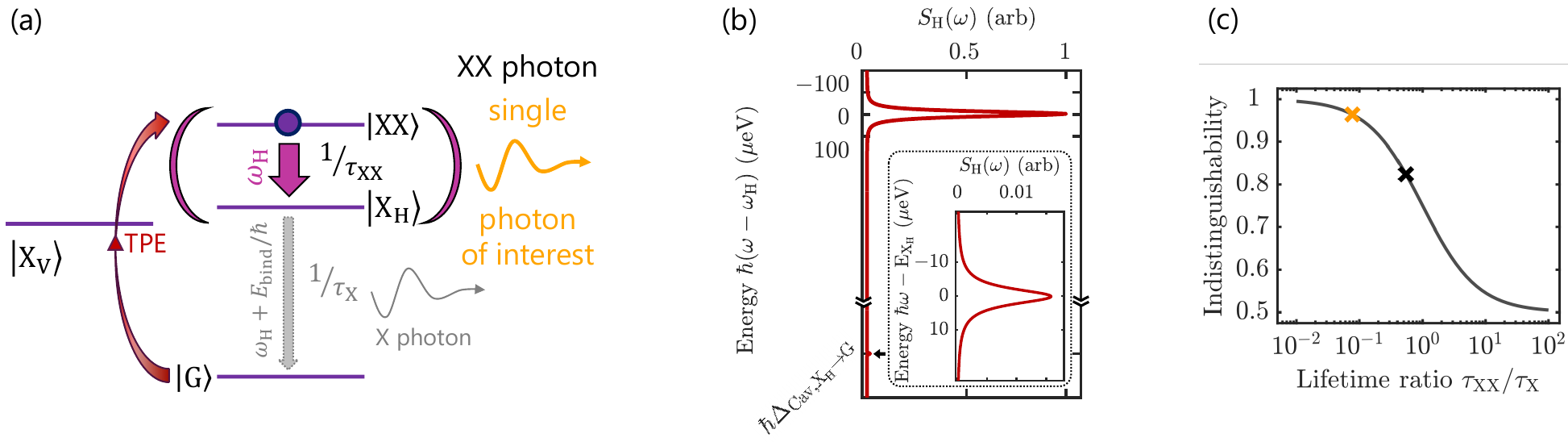} 
\caption{(a) Schematics of quantum-dot-cavity system for single-photon generation from biexciton. Considered electronic states are ground state, $\ket{\text{G}}$, two exciton states, $\ket{\text{X}_\text{H}}$ and $\ket{\text{X}_\text{V}}$, and biexciton state, $\ket{\text{XX}}$, with biexciton binding energy $\Ebind$. Electronic transitions couple to linearly polarized cavity modes with angular frequencies $\omegaH$ and $\omegaV$. The (bi-)exciton decays with rate $1/\tau_{\text{X}}$ ($1/\tau_{\text{XX}}$), which is inverse to its lifetime. Single-photon generation occurs in the H-polarization channel with H-cavity mode tuned to the XX-$\text{X}_\text{H}$ transition. The V-cavity mode (not shown) is spectrally detuned to favor emission into the H-cavity mode. In our simulations, biexciton initialization with two-photon excitation (TPE) proceeds via the V-polarization channel. Energy differences shown are not to scale. (b) Calculated cavity emission spectrum for $\Ebind=\SI{2.9}{\meV}$ and $T=\SI{4.2}{\K}$, normalized to the main cavity peak. The inset shows the side peak at $\text{X}_\text{H}$-G transition frequency. (c) Achievable photon indistinguishability for different biexciton-exciton lifetime ratios for an ideal three-level model (\Eqref{LimitOnI}, adapted from \cite{Schoell_2020,Simon_2005}). This result also applies to our Purcell-enhanced case, as explicitly derived in \Apprefs{AppX:ExpansionOfCrux}{AppX:ConnectionToExp}, and to the experimental situation with a tunable cavity \cite{Baltisberger_2025}, as further detailed in \Appref{AppX:ConnectionToExp}. The black cross marks the lifetime ratio for the naturally decaying QD and the orange cross for the Purcell-enhanced XX-X transition for the parameters of the present paper.}\label{Fig1}
\end{figure*}

In the present work we propose to avoid this complication from the outset by directly harnessing the single photon emitted from the cavity-enhanced XX-X transition as illustrated in \Fig{\ref{Fig1}}\nolinebreak (a). By using a spectrally narrow cavity on the XX-X transition \cite{Bauch_2024,Huber_2013}, we selectively extract photons from the Purcell-enhanced XX-X transition \cite{Baltisberger_2025,Ota_2011,Behrends_2026}, while suppressing emissions from the X-G transition into the cavity mode. We present a detailed theoretical exploration of this setup for single-photon generation, which includes also a first experimental measurement of single-photon generation from the biexciton, showing simultaneously high values of photon indistinguishability and single-photon purity, without the need of any signal post processing.

The present paper is organized as follows. We begin with an overview of the QD-cavity model under consideration, followed by a detailed introduction of our photon-generation scheme. Subsequently, we identify QD-cavity parameter regimes that lead to the generation of high-quality single photons, here assuming a pre-excited QD. Next we demonstrate the resilience of our approach to resonant two-photon biexciton excitation. As a benchmark for the theoretical analysis, we show good agreement of our finding with our recent experimental realization \cite{Baltisberger_2025} for a specific set of cavity and quantum dot parameters. Finally, we also briefly explore the present setup for generation of indistinguishable polarization-entangled photon pairs. In the appendices we include further analysis of results, details of system modeling, and derive essential theoretical foundations.


\section{Quantum-dot-cavity system}\label{Sec:QDCavSystem}
We model the lowest electronic excitations of a semiconductor QD as a four-level system \cite{Hanschke_2018,Ota_2011} as illustrated in \Figref{Fig1}\nolinebreak (a). This model employs typical parameters (as given below) for high-quality InGaAs QDs, which are chosen to resemble the experimental conditions with details given in \cite{Baltisberger_2025}. Phonon-related parameters are taken from \cite{Laucht_2009,Praschan_2022}. Respective parameters for phonon-mediated interactions correspond to a spherical isotropic approximation of the pancake-shaped QD as in \cite{Kaldewey_2017_Phonons}. This approximation appears to be a reasonable assumption for the purposes of this work. Parameters for phonon-induced pure dephasing are included for demonstration purposes, but overestimate such effects for our present experiments \cite{Baltisberger_2025} (see discussions below). We note that the specific choice of system parameters does not affect the primary results and conclusions. 
 
 We include the electronic ground state $\ket{\text{G}}$, two orthogonal exciton states $\ket{\text{X}_\text{H}}$ and $\ket{\text{X}_\text{V}}$, and the biexciton state $\ket{\text{XX}}$. The energies of the X$_i$, $i=\text{H,V}$, are given by $E_{\text{X}_\text{H,V}}=\EX \pm \tfrac{\Efsp}{2}$, with $\EX = \SI{1.3445}{\eV}$ and fine structure splitting $\Efsp=\SI{10.8}{\micro\eV}$. We note that the exact choice of fine structure splitting only has minimal impact on our single-photon emission results (using only the H-polarization channel for emission; see next paragraph for details). The energy of the biexciton, $E_{\text{XX}}=2\EX - \Ebind$, is set by its binding energy $\Ebind=\SI{2.9}{\meV}$ (unless specified otherwise).
 
 The electronic transitions are coupled to their respective orthogonal linearly polarized cavity modes with angular frequencies $\omega_i$, $\ i=\text{H,V}$. Here, we assume perfect alignment between the respective polarizations of the QD transition dipoles and the cavity modes (which is fulfilled to a good approximation in the experimental realization in \cite{Baltisberger_2025}). The coupling constant of electronic transitions to cavity modes is $\hbar g=\SI{20.8}{\mu\eV}$ and the photon loss is given by $\hbar\kappa=\SI{4.97}{}\cdot \hbar g \approx \SI{103.4}{\mu\eV}$, unless stated otherwise. The cavity modes are symmetrically detuned from the central frequency $\omegaC=\tfrac{1}{2}(\omegaH+\omegaV)$ with a frequency difference of $\hbar\omegaSpl=\hbar(\omegaV-\omegaH)=\SI{206.8}{\mu\eV}$. The H-polarized cavity mode is adjusted with the XX binding energy, always ensuring resonance with the XX-$\text{X}_{\text{H}}$ transition. The frequency of the V-polarized cavity mode is determined by the cavity mode splitting as given above. Additionally, the QD is coupled to a classical light field with V polarization and sufficient detuning from the cavity mode for external coherent optical driving with pulse area $\Omega_0$.
 
 System time evolution is calculated solving the polaron master equation \cite{Ota_2011,Roy_2011,Roy_2012,Gustin_2018,Roy_2012,Manson_2016,McCutcheon_2010}, which includes the coupling of the QD to its environment (full details on the modelling are given in \Appref{AppX:Theory}). This also entails radiative decay into modes other than the cavity modes, $\propto \gamma_{\text{rad}}^i$, $i=\text{X},\text{XX}$, as well as coupling to the lattice environment, with phonon-induced pure dephasing parameter, $\propto \gamma_{\text{deph}}$, and phonon-mediated interactions with the light fields, $\propto \alpha_p$. The polaron master equation approach introduces the temperature dependent rescaling factor $\AvB_T$, especially rescaling all QD-light-field couplings, i.e. $\tilde{g}=\tilde{g}_T=\AvB_T g$, $\tilde{\Omega}=\tilde{\Omega}_T=\AvB_T \Omega_0$, and $\AvB_T^2\gamma_{\text{rad}}^i$. Because we analyze different temperatures, we need to distinguish rescaled and bare parameters, especially of the QD-cavity coupling. Quantities that ultimately determine the system dynamics, i.e., the Purcell rate, but also the rates depopulating the electronic states (or, inversely, their lifetimes) refer to the rescaled parameters. Phonon-induced pure dephasing at finite temperatures is included for $T=\SI{4.2}{\K}$ (not for $T=\SI{0}{\K}$). This additional phenomenological term captures potential broadening of the zero-phonon line due to processes such as charge noise or higher-order coupling, which is known to play a role in many types of quantum dot systems \cite{Laucht_2009,Kuhlmann_2013_Noise}. We include this effect to demonstrate its potential influence, however, we note that comparison with our measurements shows that the system used in our present experiments exhibits extremely low noise and transition linewidths close to the transform limit \cite{Baltisberger_2025}. In fact, as a consequence we find very good agreement of our experimental results with the calculations for $T=\SI{0}{\K}$. We note that the dynamics induced by phonon-assisted processes change only slightly with temperature and are not much different for $T=\SI{0}{\K}$ and $T=\SI{4.2}{\K}$. Photonic correlation functions for calculation of emission spectra and quantum properties of generated photons are evaluated using the quantum regression theorem \cite{Carmichael_1999}.
 

\section{Single photons from cavity-enhanced biexciton}
The XX-X$_\text{H}$ transition is selectively enhanced with the resonant H-polarized cavity mode as introduced above \cite{Bauch_2024,Huber_2013,Ota_2011,Behrends_2026}. \FIGref{Fig1}\nolinebreak(a) shows an illustration of the emission scheme and \Figref{Fig1}\nolinebreak(b) an example of a calculated cavity emission spectrum. As highlighted in the introduction, key for highly indistinguishable photons is a small XX-X lifetime ratio (see \Figref{Fig1}\nolinebreak(c)). This originates in the need for separability between the XX and X photons emitted by the XX-X cascade, as investigated in \cite{Schoell_2020,Simon_2005}. There, based on \cite{Huang_1993}, a fundamental study on the lifetime-ratio dependence of the visibility and indistinguishability is performed, but without focusing on specific schemes for the realization of the different XX and X lifetimes. Our extensions of these studies show that cavity enhancing the XX decay in the Purcell regime ($\kappa > g \gg \gamma_{\text{rad}}$) leads to replacement of the natural XX lifetime with the inverse of the Purcell decay rate $4\tilde{g}^2/\kappa$ \cite{Tomm_2021} as detailed in \Appref{AppX:ExpansionOfCrux}. Therefore, cavity enhancing the XX can provide a lifetime ratio for which high photon indistinguishabilities are expected.

For our specific studies, we introduce the following terms. Photons emitted by the XX into the H-cavity mode are termed XX photons. Unless explicitly stated otherwise, any reference to a vectorial polarization pertains to the H channel, encompassing light modes, QD transitions, and Xs. To minimize the influence of the XX initialization process, in our simulations, the excitation is performed in the perpendicular V-polarization channel \cite{Schweickert_2018,Kuhlmann_2013}, which in the real system is eliminated in the detection by efficient polarization and moderate spectral filtering \cite{Baltisberger_2025}. To maximize emission in the H mode, the V-cavity mode is detuned as much as realistically possible; following \cite{Tomm_2021}, we choose the mode splitting as introduced above, $\hbar\omegaSpl=\hbar(\omegaV-\omegaH)=\SI{206.8}{\mu\eV}$. Ideally, the setup would eliminate any influence from the V-cavity mode by employing even larger cavity mode splitting, which may be feasible in future experiments with suitably designed cavities. Techniques described in \cite{Tomm_2021_PRApl} may further increase the mode splitting in our setup. Nonetheless, our investigation shows that the finite cavity mode splitting primarily influences the leakage of photons into the V-polarization channel, reducing emission brightness but not diminishing the quality of the single photons of interest detected in the H mode. 

While the cavity Purcell enhancement of the XX emission ensures the XX-X lifetime ratio for which high indistinguishabilities can be expected, it can also lead to additional cavity photons from the X-G transition (not included in the simplified model discussed above). Based on their origin, we call these photons X photons. As they compromise the single-photon character of the output, it is crucial to suppress their emission. For practical implementation, it is advisable to achieve this from the outset rather than resorting to additional filtering in time and/or frequency. In the following, we explore suitable regimes of QD and cavity parameters to accomplish this.


\section{General potential of the scheme} \label{Sec:GeneralPotential}
To prevent X photons from being emitted into the cavity, it is crucial to avoid resonance between the X-G transition and the cavity. In our setup, the H cavity is resonant with the XX-X transition, and thus the detuning between the X-G transition and the cavity is determined by the XX binding energy $\hbar\DeltaXHG = -\Ebind-\Efsp$. We begin by examining the influence of the XX binding energy to identify suitable QD parameters for our single-photon emission scheme from the XX. For the XX binding energy we note that changing this parameter in an experiment is not to be understood as a tunable parameter, but involves different QDs showing the respective XX binding energy. In the present section, we assume that the QD is initially in the XX state with initially empty cavity.

The quality of single photons can be characterized by the emission $\EH$, purity $\PH$ and indistinguishability $\IH$ \cite{Gustin_2020,SUPER,Gustin_2018} (definitions given in \Appref{AppX:Theory}). While the emission gives the number of photons emitted by the H-cavity mode, purity and indistinguishability quantify the single-photon characteristics of the emission. It is important to note that our analysis of quality of photons includes all photons emitted by the H cavity, not solely the XX photons. Depending on system parameters, this might include residual X photons as well. In that sense, our results give a lower bound for achievable single-photon quality. Additional spectral and/or temporal filtering \cite{Gustin_2020,Somaschi_2016,Wei_2014} is expected to enhance the photon quality further, see \Appref{AppX:SpectralFiltering} for details.

\begin{figure}
\centering
\includegraphics[width=1.0\columnwidth]{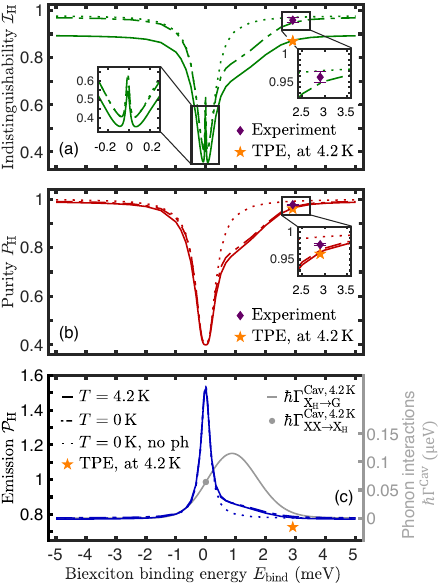} 
\caption{{Influence of biexciton binding energy and phonons on single-photon emission.} For H-mode cavity photons we show (a) photon indistinguishability $\IH$ (green), (b) single-photon purity $\PH$ (red), (c) emission $\EH$ (blue) for $T=\SI{4.2}{\K}$ (solid lines), $T=\SI{0}{\K}$ (dashed-dotted lines), and $T=\SI{0}{\K}$ without phonon interactions (dotted lines) for initially excited biexciton and empty cavity. To demonstrate insensitivity of single-photon indistinguishability and purity to resonant two-photon excitation (TPE) of the biexciton, corresponding data points are included as orange stars for pulse length $\sigma=\SI{5}{\ps}$ and pulse area $\Omega_0=\SI{5.2}{\pi}$ at $T=\SI{4.2}{\K}$ (results are insensitive to TPE pulse length; pulse parameters given in \Appref{AppX:Theory}, \TabRef{Tab1}). Results of experimental realization \cite{Baltisberger_2025} are shown as purple diamonds with error bars in (a,b). Panel (c) includes the analytical approximation for the relevant phonon-mediated QD-cavity coupling (light gray) \cite{Roy_2011} for the X-G transition, $\GammaXHG$ (solid line), and for the XX-X transition $\GammaBXH$ (single dot). Detuning between X-G transition and cavity changes with biexciton binding energy as $\hbar\DeltaXHG = -\Ebind-\Efsp \approx -\Ebind$.}\label{Fig2}
\end{figure}

\FIGref{Fig2} shows the results for H-cavity photon quality for different XX binding energies. Panel (a) shows the indistinguishability $\IH$ (green), (b) purity $\PH$ (red), and (c) emission $\EH$ (blue). The color conventions introduced here will be used consistently throughout the paper. Each panel displays results for three scenarios: $T=\SI{4.2}{\K}$ (solid lines), $T=\SI{0}{\K}$ (dashed-dotted lines), and $T=\SI{0}{\K}$ without phonons (dotted lines). We further include values with resonant TPE in the V channel (orange stars), and results of the experimental demonstration (purple diamonds with error bars in (a,b)) for the specific experimentally observed XX binding energy of $\SI{2.9}{\milli\eV}$.

At sufficiently large absolute XX binding energies, as expected for the scenario where only the XX-X transition is resonant with the cavity mode, we observe emission below unity. A maximum in emission occurs close to zero absolute XX binding energy, where also the X-G transition is near-resonant with the cavity mode. Both purity and indistinguishability exhibit similar behavior, featuring dips at zero absolute XX binding energy where the X emission degrades the single-photon character of the emission. Purity nearly reaches the value for perfect single photons at sufficiently large absolute XX binding energies that exceed the cavity width. When phonon-mediated QD-cavity interactions are considered, we observe more pronounced dips (and corresponding peaks in emission) with a pronounced asymmetry around zero XX binding energy with a broader dip on the positive XX binding energy side (e.g., induced by phonon emission processes). Unlike purity, indistinguishability shows two dips around zero XX binding energy and a slight increase in between them, forming a small local indistinguishability maximum.

The behavior of purity and emission can be explained as follows: at larger absolute XX binding energies, the cavity emits only XX photons. As in that case the cavity is off-resonant to the X-G transition, X photons are primarily emitted through non-cavity modes, leading to emission levels below unity and the near-ideal purity. This also explains the observed high indistinguishability. The fact that indistinguishability, in contrast to purity, still falls short of its ideal value originates in the fundamental limitation imposed by the lifetime ratio for the current QD-cavity coupling and cavity loss. The emission below unity in the H mode is due to XX photons being emitted into non-cavity modes and in the V mode. Near $\Ebind=\SI{0}{\milli\eV}$, resonance with the X-G transition can be clearly observed: emission reaches a maximum while purity drops to a minimum. This behavior results from additional X photons being emitted into the cavity, which compromise the single-photon nature and also degrade indistinguishability. The two dips in indistinguishability arise for the following reasons: at exact resonance, all photons share the same frequency, resulting in slightly increased indistinguishability. At small detunings, X photons exhibit slightly different frequencies, leading to further decreases in indistinguishability and creation of two observable minima. The more pronounced dips (or peaks) towards small positive XX binding energies for $T=\SI{0}{\K}$ and $T=\SI{4.2}{\K}$ can be attributed to the inclusion of phonon-mediated QD-cavity interactions. Phonon-mediated cavity feeding at low temperatures occurs only when transition energies exceed the cavity frequency (phonon emission) \cite{Roy_2011,Hughes_2013}, resulting in increased emission and consequently decreased purity and indistinguishability in this XX binding energy regime. For visualization, the light gray line in panel (c) for $T=\SI{4.2}{\K}$ represents the analytic coupling rate $\GammaXHG$ given in \Eqref{PhononIndQDCavCoupling}, approximating phonon interactions with a Lindbladian. The slightly more pronounced effects at $T=\SI{4.2}{\K}$ compared to $T=\SI{0}{\K}$ are due to the increase in phonon-mediated QD-cavity interactions with increased temperature. The noticeable reduction in indistinguishability at $T=\SI{4.2}{\K}$ compared to the other cases, is explained by the additional pure dephasing. Finally, we comment on the not entirely symmetric behavior of the photon quality quantifiers at small absolute XX binding energies. This occurs for all photon quality quantifiers and even in the phonon-free case and is rooted in the finite fine structure splitting, which breaks the symmetry in the detuning between the X-G transition and the cavity.

Our results indicate that the proposed scheme, in which the XX lifetime is selectively reduced, is capable of generating high-quality single photons, provided that the XX binding energy is sufficiently large. This is also confirmed by the experimental demonstration \cite{Baltisberger_2025}, which uses a QD with $\Ebind=\SI{2.9}{\meV}$. The measured visibility $\curlyV=0.94 \pm 0.02$ and $\gtwo=0.023 \pm 0.002$ of the XX photon, converted into measures used thus far are included in \Figref{Fig2} as purple diamonds; details are given in \Apprefs{AppX:ExpansionOfCrux}{AppX:ConnectionToExp}. Excellent agreement of experiment and theory is found. A slight difference occurs for the indistinguishability in \Figref{Fig2}\nolinebreak (a): the measurement value is quite well reproduced by our calculations, but for the pure-dephasing-free case of $T=\SI{0}{\K}$. The value of the case for $T=\SI{4.2}{\K}$ which includes pure dephasing with $\SI{4.2}{\mu\eV}$ is lower. This is because the source used in our present experiment exhibits low noise and transitions operate close to the transform limit \cite{Baltisberger_2025}. For our in-depth theoretical analysis we continue to include pure dephasing as this delivers additional insights into the potential influence of the different processes that may play an important role in other physical realizations of the proposed approach.

\begin{figure}
\centering
\includegraphics[width=1.0\columnwidth]{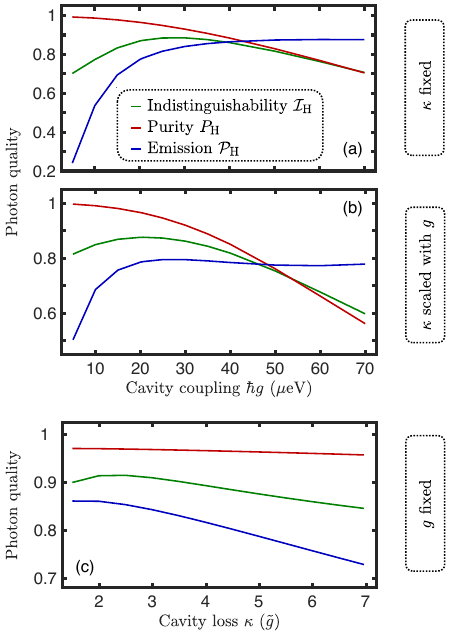} 
\caption{{Influence of QD-cavity coupling and cavity loss on single-photon emission.} Shown are photon indistinguishability $\IH$ (green), single-photon purity $\PH$ (red), and emission $\EH$ (blue) at $T=\SI{4.2}{\K}$ and $\Ebind=\SI{2.9}{\meV}$. (a,b) Influence of QD-cavity coupling for (a) fixed cavity loss $\kappa=\SI{4.97}{}\cdot \left. g\right|_{\hbar g=\SI{20.8}{\mu\eV}}$ and (b) cavity loss that scales with QD-cavity coupling as $\kappa=\SI{4.97}{}\cdot\tilde{g}_{T}$ ($\tilde{g}=\tilde{g}_T=\AvB_T g$). (c) Influence of cavity loss for fixed QD-cavity coupling $\hbar g=\SI{20.8}{\mu\eV}$. Results shown are for initially excited biexciton and empty cavity. Corresponding results for $T=\SI{0}{\K}$ are given in \Appref{AppX:DiffTemp}.} \label{Fig3}
\end{figure}

Having analyzed the influence of the XX binding energy above, now we fix it at the experimental value of $\SI{2.9}{\meV}$ and continue to analyze the influence of the cavity parameters. We focus on the case of $T=\SI{4.2}{\K}$; results for the other two cases discussed above are given in \Appref{AppX:DiffTemp}. \FIGref{Fig3}\nolinebreak (a) and (b) show the results for different QD-cavity couplings $g$ for (a) fixed cavity loss $\kappa=\SI{4.97}{}\cdot \left. g\right|_{\hbar g=\SI{20.8}{\mu\eV}}$, and (b) cavity loss that scales with QD-cavity coupling as $\kappa=\SI{4.97}{}\cdot\tilde{g}_{T}$ ($\tilde{g}=\tilde{g}_T=\AvB_T g$). In \Figref{Fig3}\nolinebreak (c) we fix the QD-cavity coupling ($\hbar g=\SI{20.8}{\mu\eV}$) and show the results for different cavity losses $\kappa$. Similar trends can be observed for the photon emission. Increasing $g$ (decreasing $\kappa$) increases the generation of photons in the cavity and thereby the emission. In the weak coupling limit of the Purcell regime, this is directly explained by the Purcell decay rate $4\tilde{g}^2/\kappa$. For stronger QD-cavity coupling, X photons that are emitted into the cavity lead to a further increase of the emission. This holds true for the X photons emitted off-resonantly into the cavity, but more importantly also for the photons generated by phonon-assisted cavity feeding. While resonant QD-cavity coupling scales with $g$ (for the XX photons), phonon-assisted cavity feeding scales with $g^2$, see \Eqref{PhononIndQDCavCoupling}. The increasing number of X photons emitted into the cavity explains immediately the visible decrease of the purity for larger QD-cavity coupling. In contrast to that, the purity decreases very slightly with larger cavity losses (almost constant with only a $\SI{1}{\percent}$ change for the parameter range shown, see \Figref{Fig3}\nolinebreak(c)). This can be explained by the broader cavity resonance, which increases the off-resonant emission of X photons into the cavity, while phonon-mediated QD-cavity coupling does not change as $g$ is constant. Finally, the indistinguishability reaches a maximum and then decreases when further increasing $g$ or $\kappa$ (see \Figref{Fig3}\nolinebreak(a) and (b)). Therefore, the indistinguishability shows similar behavior as the purity for larger $g$, which is explained analogously. However, while the purity converges towards unity for smaller probability of X photon emission into the cavity (low $g$ or $\kappa$), the indistinguishability decreases after reaching its maximum. For decreasing $g$ this has two reasons. Decreasing $g$ decreases the Purcell rate and thereby increases the XX-X lifetime ratio. Therefore, the maximally achievable indistinguishability is reduced. If pure dephasing is included, as in \Figref{Fig3}, this further decreases the indistinguishability. For fixed dephasing, this becomes more relevant for smaller $g$ as the QD decays more slowly so that dephasing has more impact. Finally, compared to the stronger sensitivity to the QD-cavity coupling $g$, cavity loss has a relatively minor influence on photon quality measures, particularly regarding the single-photon character (see \Figref{Fig3}\nolinebreak(c)).

In summary, we clearly identify parameter regimes where cavity-induced selective lifetime reduction of the XX state leads to high-quality single photons emitted from the XX state, including an experimental demonstration for one specific parameter set in an existing device.

Next we discuss the cavity emission spectra generated by our scheme. \FIGref{Fig1}\nolinebreak(b) shows an example of a cavity emission spectrum for $T=\SI{4.2}{\K}$. A pronounced peak is observed at the cavity frequency, with a distinct smaller peak appearing at the X-G transition frequency. Note that we show the emission spectrum of the cavity, not of the QD. Therefore, only QD emission into the cavity mode is included. Apart from cavity resonant emission of the XX photons, this includes phonon-mediated photons emitted into the cavity mode (phonon-assisted cavity feeding) as well as X photons emitted off-resonantly into the cavity. The separation between the cavity and X peaks suggests that spectral filtering \cite{Gustin_2020,Somaschi_2016,Wei_2014} at the cavity frequency can be used to improve photon quality (as shown in \Appref{AppX:SpectralFiltering}). Thereby, one source of reduced single-photon purity (and therefore also indistinguishability) of the cavity emission is eliminated. For elimination of X photons emitted via phonon-assisted cavity feeding, temporal filtering may be a viable solution. This is supported by the time-resolved spectral analysis of the emission (see \Appref{AppX:SpectralFiltering}, \Figref{FigA5}), indicating that the X photons are emitted on much longer time scales (a few hundred picoseconds). The XX photons are emitted after tens of picoseconds, indicating that these photons can be separated in time from the major part of the photons emitted from the X. Ideally both filtering mechanisms are used in moderation only to leave the crucial part of the emission (the XX-cavity photons) intact. We stress that spectral filtering cannot serve to alleviate the physical limit set for the indistinguishability by the X-XX lifetime ratio, nor eliminate the pure dephasing.

To quantify the potential performance of our single-photon generation scheme and for comparison with other single-photon sources we report here the relevant performance quantifiers. For XX binding energy $\Ebind=\SI{5}{\meV}$ at $T=\SI{0}{\K}$ with moderate spectral filtering (filter bandwidth $\muS=\kappa$) we obtain $\gtwo=0.0041$ and $\curlyV=0.9607$ (parameters as in our present experiment, only XX binding energy is slightly higher). This is the highest value we observe in our current simulations and it is obtained without any optimization of QD-cavity coupling or cavity loss. The case $T=\SI{0}{\K}$ is chosen here for this initial benchmarking as it appears to best approximate the present experiment (see \Figref{Fig2} and \Figref{FigA4}) with low noise and transitions close to the transform limit \cite{Baltisberger_2025}. Next, we briefly compare our results with other (single-) photon sources. In \cite{Liu_2019}, both transitions are enhanced in a bullseye cavity. Here, our scheme of selectively enhancing only the XX-X transition has an advantage as it optimizes single-photon indistinguishability. Compared with resonantly excited single‑photon sources \cite{Somaschi_2016,Wang_2016,Tomm_2021,Ding_2025,Liu_2018}, our scheme exploits the advantage of a three‑level system and we demonstrate that the achievable $\gtwo$ and the simultaneously efficient photon extraction can compete with those sources. The HOM visibilities obtained with our approach will benefit from further parameter optimization, with our theoretical calculations guiding this process for the future.

Overall, our scheme possesses clear advantages that lie in its simplicity and the absence of any need for aggressive emission post‑processing. With targeted device optimization, this approach may ultimately outperform existing sources.
 

\section{Performance with two-photon excitation}\label{Sec:TPE}
After having demonstrated that our scheme yields high-quality single photons, we continue by investigating the influence of the XX excitation process. The optical XX initialization can affect photon generation and quality of generated photons \cite{Hanschke_2018,Heinisch_2024,Vannucci_2023}. Here these effects can be minimized by XX excitation with a V-polarized laser polarized perpendicular to the H-cavity mode \cite{Schweickert_2018,Kuhlmann_2013} as used in our simulations. For different excitation polarization, spectral filtering can be used to further improve the single-photon character (see \Appref{AppX:SpectralFiltering}). However, since this would disallow direct comparison with previously presented results (unfiltered), we continue with perpendicular polarizations for XX initialization and photon generation. Furthermore, the polarization-distinguishing configuration also leads to single-photon generation in the H mode without any photon number coherence \cite{Hagen_2025} (not shown, but verified numerically). Given that resonant TPE is a relatively easy to implement and widely used XX excitation method \cite{Schweickert_2018}, it serves as a suitable candidate to test our single-photon generation scheme. 

We observe that the single-photon character of the H-polarized cavity photons is largely unaffected by TPE in the V mode. We checked this explicitly for various pulse lengths for the XX binding energy of our experiment ($\SI{2.9}{\meV}$) and as well for smaller and larger XX binding energies. Required pulse areas were determined through parameter sweeps leading to perfect XX initialization in the loss-free case; resulting pulse parameters for $E_{\text{bind}}=\SI{2.9}{\meV}$ are given in \Appref{AppX:Theory}, \TabRef{Tab1}. We note here explicitly that the pulse areas, i.e., the pulse area stated below, must be divided by $\AvB_T$ to compensate for the rescaling in the polaron master equation. For sufficiently large XX binding energies we observe nearly the same high single-photon quality as for the initially excited XX discussed above; see results in \Figref{Fig2} for pulse length $\sigma=\SI{5}{\ps}$ and pulse area $\Omega_0=\SI{5.2}{\pi}$. The robustness of the single-photon character can be attributed to the use of perpendicular polarization channels for QD excitation and photon generation \cite{Tomm_2021}. The observed emission is typically five to ten percent lower than that of the initially excited XX, which is mostly due to the imperfect initialization of the XX through TPE in the presence of losses. With the observed robustness of our single-photon generation method, it can also be successfully combined with other coherent excitation methods in the opposite polarization channel, e.g., the SUPER scheme, which is expected to eliminate the observed reduction in the emission \cite{Heinisch_2024} but may induce additional noise, or with rapid adiabatic passage \cite{Kaldewey_2017}. The (minor) aspects discussed that cause emission below unity in our scheme are not rooted in fundamental limits of the approach but can be removed by further optimization. The intrinsic problems that plague resonant single-photon generation from two-level systems -- namely quantum-emitter re-excitation and difficult (spectral) separation of excitation laser from generated single photons \cite{GonzalezRuiz_2025,Hanschke_2018,Mueller_2015,Fischer_2017,Fischer_2_2017,Heinisch_2024} -- play no role in our approach.


\section{Entangled photon pairs}\label{Sec:PhotonPairs}
In this section, we briefly explore the potential of our setup to produce high-quality indistinguishable polarization-entangled photon pairs. In addition to the XX photons this also requires a high indistinguishability for the X photons. Following our investigations in \Apprefs{AppX:ExpansionOfCrux}{AppX:ConnectionToExp}, our scheme is expected to be capable of generating such photons. We refer to \Figref{Fig1}\nolinebreak (c), which applies not only to photons from the XX but also to photons from the X. Initial calculations for the phonon-free scenario (with indistinguishability of about $\SI{98}{\percent}$), along with experimental results \cite{Baltisberger_2025}, confirm this. The other requirement is high polarization entanglement, which can be measured by the concurrence $\curlyC$ \cite{Wooters_1998,Cygorek_2018,Heinze_2017,Seidelmann_2021,Liu_2025}. However, our initial calculations at $\SI{0}{\K}$ without taking phonon interactions into account reveal that our current setup does not support this ($\curlyC=0.13$). This is because of the which-path information introduced by our setup, which is designed to separate QD excitation and photon emission. Favoring high emission in the H mode, the V-cavity mode is detuned in frequency. Therefore, entanglement is spoiled \cite{Hudson_2007,Cygorek_2018,Seidelmann_2022,Seidelmann_2021}. Nevertheless, with adjustments that eliminate the polarization splittings (zero cavity mode splitting and zero fine structure splitting), we observe perfect entanglement with $\curlyC = 1 $. Consequently, the generation of high-quality entangled photon pairs, with simultaneously high indistinguishability, is feasible with minor adjustments to our setup. Promising investigations on that using a bullseye cavity and TPE were done for example in \cite{Liu_2019}. However, QD excitation might necessitate more complex excitation methods \cite{Heinisch_2024, Bracht_2023,Vannucci_2023} or filtering, as TPE is known to undermine entanglement \cite{Seidelmann_2022, Heinisch_2024}. 


\section{Conclusions}
We have proposed and demonstrated a scheme for generating high-quality single photons from the cavity-enhanced biexciton-to-exciton transition in a semiconductor quantum dot. By design, this approach eliminates emitter re-excitation (optimizing single-photon purity) and enables straightforward separation of emitted photons from the excitation laser. Through combined theoretical analysis and first experimental measurements on a quantum dot inside in a high-quality open microcavity, we report single photons with high indistinguishability, overcoming the limitations set by the intrinsic biexciton-exciton lifetime ratio.

With resonant two-photon excitation, the best performance is achieved when biexciton initialization and photon generation occur in orthogonal polarization channels. Guided by the system parameters of the quantum-dot-cavity system used in our experiments, with sufficiently large biexciton binding energy and very moderate spectral filtering applied, $\gtwo=0.0041$ and $\curlyV=\SI{96.07}{\percent}$ are readily achieved. In our present experiments with a device that has not been specifically engineered as a biexciton-based single-photon source, $\gtwo=0.023 \pm 0.002$ and $\curlyV=94 \pm \SI{2}{\percent}$ are achieved with emission detected as is (no post processing).

Our approach effectively resolves both re-excitation and photon-extraction challenges, with future targeted engineering potentially leading to a deterministic single-photon source outperforming conventional resonant excitation schemes. To achieve this goal, emphasis should be placed on reducing cavity loss $\kappa$, and, more critically, on further suppressing exciton decay. Both would further decrease the biexciton-exciton lifetime ratio, thereby enhancing single-photon visibility. In addition to further advancing the strategies analyzed in the present work -- namely using high biexciton binding energies and spectral filtering -- temporal filtering of the rapidly emitted cavity-enhanced single photons may be considered. Furthermore, switching from InGaAs to GaAs quantum dot systems is expected to further enhance photon quality by reducing phonon interactions and offering larger biexciton binding energies \cite{Zhai_2020,Zhai_2022,Liu_2019,BassoBasset_2017}.

While putting the finishing touches to the present manuscript, another experimental demonstration of cavity-enabled lifetime engineering in the biexciton-exciton cascade was reported for telecommunication wavelengths \cite{Behrends_2026}, further underlining the importance of our in-depth analysis of the presented approach.


\begin{acknowledgments}
N.H. thanks Nikolas K{\"o}cher for various fruitful discussions. The work in Paderborn was supported by the Deutsche Forschungsgemeinschaft (German Research Foundation) through the transregional collaborative research center TRR~142/3-2022 (231447078, project C09) and by Paderborn Center for Parallel Computing, PC$^2$. This work is further supported by the German Federal Ministry of Research, Technology and Space (BMFTR) through the project QR.N (16KIS2206). This project has received funding from the European Research Council (ERC) under the European Union’s Horizon 2020 research and innovation program (LiNQs, grant agreement 101042672). K.D.J. and S.S. acknowledge funding from the Ministry of Culture and Science of North Rhine-Westphalia for the Institute for Photonic Quantum Systems (PhoQS). The work in Basel was funded by Swiss National Science Foundation project 200020\_204069. S.R.V. and A.L. gratefully acknowledge funding by the EUROSTARS project QTRAIN under BMFTR grant 13N17328, the QuantERA project under BMFTR grant 16KIS2061, and QR.N under BMFTR grant 16KIS2200.
\end{acknowledgments}


\appendix


\section{Results for different temperatures}\label{AppX:DiffTemp}

\begin{figure}
\centering
\includegraphics[width=0.8\columnwidth]{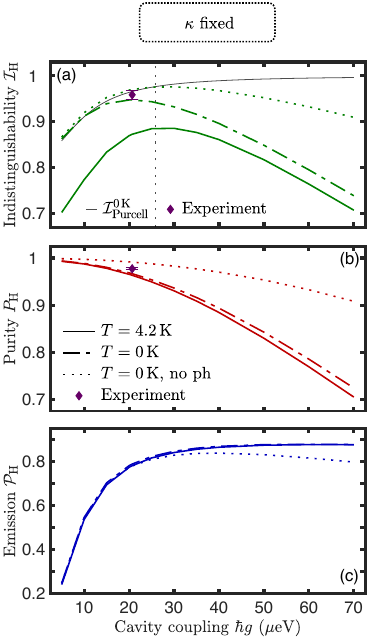} 
\caption{{Influence of QD-cavity coupling on single-photon emission for different temperatures and phonon models for fixed cavity loss.} Presented is an extended and rearranged version of \Figref{Fig3}\nolinebreak (a). Shown are (a) indistinguishability $\IH$ (green), (b) purity $\PH$ (red), and (c) emission $\EH$ (blue) at $T=\SI{4.2}{\K}$ (solid lines), $T=\SI{0}{\K}$ (dashed-dotted lines), and $T=\SI{0}{\K}$ without phonons (dotted lines) for $\Ebind=\SI{2.9}{\meV}$. Panel (a) includes the analytical approximation of the indistinguishability $\curlyI_{{\rm Purcell}}^{0\,{\rm K}}$ at $T=\SI{0}{\K}$ (dark gray line, \Eqref{LimitOnI} with \Eqref{Eq:FullDecayRates}), valid for weak coupling ($g \ll \kappa/4$, black dashed vertical line) in the Purcell regime, as clarified in \Apprefs{AppX:ExpansionOfCrux}{AppX:ConnectionToExp}. Results for fixed cavity loss $\kappa=\SI{4.97}{}\cdot \left. g\right|_{\hbar g=\SI{20.8}{\mu\eV}}$ and initially excited biexciton and empty cavity. Results of experimental realization \cite{Baltisberger_2025} are shown as purple diamonds with error bars in (a,b).}\label{FigA1}
\end{figure}

\begin{figure}
\centering
\includegraphics[width=0.8\columnwidth]{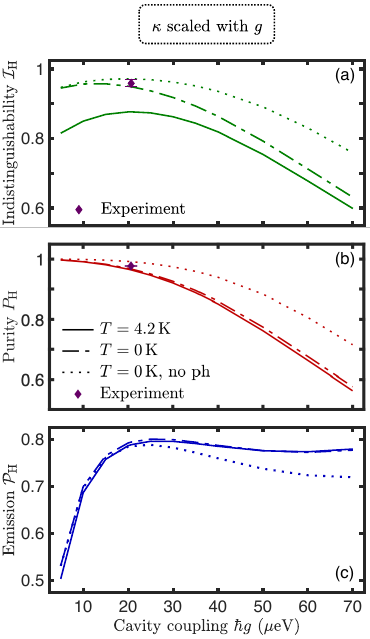} 
\caption{{Influence of QD-cavity coupling on single-photon emission for different temperatures and phonon models for scaled cavity loss.} Presented is an extended and rearranged version of \Figref{Fig3}\nolinebreak (b). Shown are (a) indistinguishability $\IH$ (green), (b) purity $\PH$ (red), and (c) emission $\EH$ (blue) at $T=\SI{4.2}{\K}$ (solid lines), $T=\SI{0}{\K}$ (dashed-dotted lines), and $T=\SI{0}{\K}$ without phonons (dotted lines) for $\Ebind=\SI{2.9}{\meV}$. Cavity loss scales with QD-cavity coupling, as $\kappa=\SI{4.97}{}\cdot\tilde{g}_T$ ($\tilde{g}=\tilde{g}_T=\AvB_T g$). Results for initially excited biexciton and empty cavity. Results of experimental realization \cite{Baltisberger_2025} are shown as purple diamonds with error bars in (a,b).}\label{FigA2}
\end{figure}

\begin{figure}
\centering
\includegraphics[width=0.8\columnwidth]{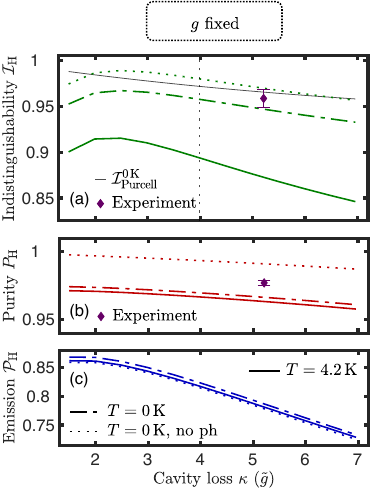} 
\caption{{Influence of cavity loss on single-photon emission for different temperatures and phonon models.} Presented is an extended and rearranged version of \Figref{Fig3}\nolinebreak (c). Shown are (a) indistinguishability $\IH$ (green), (b) purity $\PH$ (red), and (c) emission $\EH$ (blue) at $T=\SI{4.2}{\K}$ (solid lines), $T=\SI{0}{\K}$ (dashed-dotted lines), and $T=\SI{0}{\K}$ without phonon interactions (dotted lines) for $\Ebind=\SI{2.9}{\meV}$. Panel (a) includes the analytical approximation of the indistinguishability $\curlyI_{{\rm Purcell}}^{0\,{\rm K}}$ at $T=\SI{0}{\K}$ (dark gray line, \Eqref{LimitOnI} with \Eqref{Eq:FullDecayRates}), valid for weak coupling ($\kappa \gg 4g$, black dashed vertical line) in the Purcell regime, as clarified in \Apprefs{AppX:ExpansionOfCrux}{AppX:ConnectionToExp}. Results for fixed QD-cavity coupling $\hbar g=\SI{20.8}{\mu\eV}$ ($\tilde{g}=\tilde{g}_T=\AvB_T g$). Results for initially excited biexciton and empty cavity. Results of experimental realization \cite{Baltisberger_2025} are shown as purple diamonds with error bars in (a,b).}\label{FigA3}
\end{figure}

In this section, we compare the results of our numerical investigations on single-photon qualities across different temperatures and phonon influences. This complements the findings presented in \Figref{Fig3} regarding the cases of $T=\SI{0}{\K}$ with and without phonon interactions. Again, we present results for the binding energy $\Ebind=\SI{2.9}{\meV}$. For our investigations into different parameters, we follow the same order as in the main text. First, we examine the influence of the QD-cavity coupling $g$. \FIGref{FigA1} shows results for fixed cavity loss, while \Figref{FigA2} presents results where the cavity loss $\kappa$ is scaled with $g$: $\kappa=\SI{4.97}{}\cdot\tilde{g}$. After that, we explore the influence of cavity loss $\kappa$ in \Figref{FigA3}. The figures are organized similarly to \Figref{Fig2}. Hence, panel (a) shows indistinguishability $\IH$ in green, panel (b) purity $\PH$ in red, and panel (c) emission $\EH$ in blue. Each panel depicts results for three temperature scenarios: $T=\SI{4.2}{\K}$ (solid lines), $T=\SI{0}{\K}$ (dashed-dotted lines), and $T=\SI{0}{\K}$ without phonon interactions (dotted lines). Results of the experimental realization \cite{Baltisberger_2025} are shown as purple diamonds with error bars in (a,b). In \Figref{FigA1} and \Figref{FigA3}, the panels (a) further include the analytical results for the indistinguishability $\curlyI_{{\rm Purcell}}^{0\,{\rm K}}$ at $T=\SI{0}{\K}$ (dark gray lines, \Eqref{LimitOnI} with \Eqref{Eq:FullDecayRates}), valid for weak coupling ($\kappa \gg 4g$, indicated by the black dashed vertical lines) in the Purcell regime (see \Apprefs{AppX:ExpansionOfCrux}{AppX:ConnectionToExp} for details).

Here, we focus on important observations and differences from the results already presented in the main text. A key feature of \Figref{FigA1} and \Figref{FigA2} is the indistinguishability shown in panel (a), especially for the cases of $T=\SI{0}{\K}$. For larger QD-cavity coupling, that case approaches the scenario of $T=\SI{4.2}{\K}$, while it tends toward the phonon-free case as $g$ decreases. This can be explained by the interplay of two phonon-related effects that vary with different $g$, temperatures, and different phonon influences: phonon-induced pure dephasing and phonon-mediated QD-cavity interactions. While the former is present only at finite temperatures -- specifically $T=\SI{4.2}{\K}$ in our case -- the latter occurs at all temperatures and increases with $g$, while vanishing at small $g$. Therefore, the relative impact of these processes change from dephasing only to domination of phonon-mediated QD-cavity interactions as $g$ increases. Further, phonon-mediated QD-cavity interactions are slightly increased at $T=\SI{4.2}{\K}$ due to the higher temperature. In the phonon-free version at $T=\SI{0}{\K}$, neither phonon-mediated QD-cavity interactions nor phonon-induced pure dephasing are present. For the $T=\SI{0}{\K}$ case the aforementioned points explain why this case approaches its phonon-free version at small $g$. Generally, no pure dephasing occurs for $T=0$ in our model and phonon-mediated QD-cavity interactions are minimal due to small $g$. Increasing the QD-cavity coupling also increases phonon-mediated QD-cavity interactions, causing the $T=\SI{0}{\K}$ case to perform worse than the phonon-free case. For $T=\SI{4.2}{\K}$, phonon-mediated QD-cavity coupling dominates over pure dephasing. The $T=\SI{0}{\K}$ case approaches this situation at larger $g$, differing only slightly in phonon-mediated QD-cavity interactions and the residual impact of dephasing.

The effect of increased phonon-mediated QD-cavity interactions with growing QD-cavity coupling is also visible in the purity and in the emission (\Figref{FigA1} and \Figref{FigA2}, panels (b) and (c), respectively). Unlike indistinguishability, these quantities are insensitive to pure dephasing. Therefore, all three cases converge to the same value when $g$ decreases, where phonon-mediated QD-cavity interactions are reduced. \FIGref{FigA3} reveals the influence of the cavity loss $\kappa$, although no new insights emerge. The different situations respectively differ consistently in the impact of phonon-mediated QD-cavity interactions and phonon-induced pure dephasing across the swept parameters, consistent with the observations and interpretations provided in the main text for \Figref{Fig2}.

Finally we comment on the analytic results included in \Figref{FigA1}\nolinebreak(a) and \Figref{FigA3}\nolinebreak(a) as dark gray lines. These results are obtained taking into account the overall decay rates for both transitions, \Eqref{Eq:FullDecayRates}, i.e., (Lorentz-masked) Purcell rate plus radiative decay rate, and assuming $\gtwo=0$ (i.e., no influence of X photons on indistinguishability). Good agreement between the analytic approximation and the numerical simulations is observed in the validy range of the analytics, that is, in the weak coupling limit ($\kappa \gg 4g$) of the Purcell regime. There, deviations from the numerical simulations are on the order of only $\SI{1}{\percent}$. Larger deviations at larger QD-cavity couplings (smaller cavity losses) as visible in \Figref{FigA1} (\Figref{FigA3}) can be explained with the violation of the weak coupling condition, as in these cases $g \nless \kappa/4$. We do not include the analytic approximation in \Figref{FigA2} as with $\kappa$ decreasing with $g$, the analytic approximation $\kappa \gg \gamma_\text{rad}^{\text{XX,X}}$ would be violated (see \Appref{AppX:ExpansionOfCrux} for details).


\section{Theory}\label{AppX:Theory}
In this section, we present the theory used for our investigations. The full Hamiltonian in the lab frame is given by
\begin{equation}
    H \ = \ H_\text{QD} \ + \ H_\text{cav} \ + \ H_\text{QD-cav} \ + \ H_\text{QD-light},
\end{equation}
containing the free evolution of the QD \cite{Ota_2011},
\begin{equation}
    H_\text{QD} \ = \ \sum_{i=\text{G},\text{X}_\text{H,V},\text{XX}} E_i \ketbra{i}{i},
\end{equation}
with the respective QD energies $E_i,\ i=\text{G},\text{X}_\text{H,V},\text{XX}$, and the free evolution of the cavity,
\begin{equation}
    H_\text{cav} \ = \ \hbar \sum_{i=\text{H,V}} \omega_{i} a_i^\dagger a_i,
\end{equation}
with cavity mode creation and annihilation operators (CAOs) $a_i^{(\dagger)}$ and angular frequencies $\omega_i$, with $i=\text{H,V}$. Both interaction terms of the QD with either classical or quantized cavity light fields are treated in dipole and rotating wave approximation. The interaction Hamiltonian between QD and cavity reads
\begin{equation}
    H_\text{QD-cav} \ = \ \hbar g \sum_{i=\text{H,V}} \big(a_i\ketbra{\text{X}_i}{\text{G}} + a_i \ketbra{\text{XX}}{\text{X}_i}\big) \ + \ \hc 
\end{equation}
with real-valued coupling constant $g$. Therefore, we use the same coupling for both cavity modes. As only H polarization is measured in the experiment and our V-cavity mode is detuned from all QD transitions, slight differences between the couplings as they might appear in the experiment should be without significant impact on our theory results. In general, only the emission can be impacted by that, but not the single-photon character of the H-cavity mode. The interaction Hamiltonian between QD and driving field (V polarized) is given as
\begin{equation}
    H_\text{QD-light} = -\frac{\hbar}{2} \Omega(t) \big( \ketbra{\text{X}_\text{V}}{\text{G}} + \ketbra{\text{XX}}{\text{X}_\text{V}}\big) + \hc.
\end{equation}
This model implies that the classical light field propagates in an additional mode and neglects direct laser driving of the V-cavity mode. For the TPE we use Gaussian pulses,
\begin{equation}\label{FieldAppendix}
\begin{aligned}
\Omega(t) \ &= \ \frac{\Omega_0}{\sqrt{2 \pi \sigma^2}} \, \E^{-(t-t_0)^2/(2\sigma^2)} \, \E^{-\I \omega (t-t_0)},
\end{aligned}
\end{equation}
with pulse area $\Omega_0$, duration $\sigma$ and energy $\hbar\omega=E_\text{XX}/2$. The time shift $t_0$ is chosen such that the pulse emerges after $t=0$. Parameters used can be found in \TabRef{Tab1}.
\begin{table}
    \centering
    \begin{tabularx}{\columnwidth}{*{5}{c}}
     \hline\hline
      $ \ \ \ \quad \quad $ & $ \ \ \quad \quad $ & $ \ \ \quad \quad $ & $\sigma$ $(\SI{}{\pico\s}) \ $ & $ \ \Omega_0$ $(\SI{}{\pi}) \ \ $ \\\hline
   & & & $3$ & $4.2$ \\
   & & & $5$ & $5.2$ \\
   & & & $6$ & $5.7$ \\
   & & & $9$ & $6.9$ \\
   & & & $12$ & $7.8$ \\
   & & & $15$ & $8.6$ \\\hline
    \end{tabularx}
    \caption{Pulses for two-photon excitation that lead to $\pi$ excitation of biexciton for the biexciton binding energy $\Ebind=\SI{2.9}{\meV}$ in the loss-free case, i.e., for $\AvB=1$ (without rescaling). Pulse areas must be divided by $\AvB$ to compensate for the rescaling in the polaron master equation and maintain $\pi$ excitation.}\label{Tab1}
\end{table}
The equation of motion for the system density operator uses the polaron master equation model \cite{Roy_2012,Roy_2011,Gustin_2018,Roy_2012,Manson_2016,McCutcheon_2010,Praschan_2022} 
\begin{equation}\label{vNeqAppendix}
    \dot{\rho} \ = \ - \tfrac{\I}{\hbar} \, \big[ H , \rho \big] \ + \ \curlyL_\text{cav} \, \rho \ + \ \curlyL_\text{rad} \, \rho \ + \ \curlyL_\text{phonons} \, \rho .
\end{equation}
The QD-light coupling introduced beforehand needs to be rescaled by the averaged phonon displacement operator
\begin{equation}
    \AvB \ = \ \AvB_T \ = \ \exp \left[ -\frac{1}{2}\int_0^\infty\D\omega \frac{J(\omega)}{\omega^2}\coth\left(\frac{\hbar \omega}{2 k_{\text{B}}T}\right)\right];
\end{equation}
$\tilde{g}=\tilde{g}_T=\AvB_T g$, $\tilde{\Omega}=\tilde{\Omega}_T=\AvB_T \Omega_0$. Here, the phonon bath dynamics are introduced by their spectral distribution
\begin{equation}
    J(\omega) \ = \ \alpha_p \omega^3 \exp \left[ - \frac{\omega^2}{2 \omega_b^2} \right],
\end{equation}
where $\alpha_p=\SI{0.03}{\square\ps}$ is the phonon coupling factor, $\hbar\omega_b=\SI{1}{\meV}$ the phonon cutoff energy, $T$ the temperature and $k_{\text{B}}$ the Boltzmann constant \cite{Praschan_2022}. These values are obtained when approximating the pancake-shaped QD in \cite{Kaldewey_2017_Phonons} as a spherical isotropic QD, which appears to be a reasonable assumption for the purposes of the present work. \Eqref{vNeqAppendix} contains further the super operators 
\begin{equation}\label{Eq:CavLindbladian}
    \curlyL_\text{cav} \, \rho \ = \ \kappa \ \sum_{i=\text{H,V}} \curlyL_{a_i} \, \rho, 
\end{equation}
\begin{equation}
    \curlyL_\text{rad} \, \rho \ = \ \AvB^2\ \sum_{{\hat{\sigma}}} \gamma_\text{rad}^{\hat{\sigma}} \curlyL_{\hat{\sigma}} \, \rho,\label{RadDec}
\end{equation} 
describing cavity loss with the rate $\kappa$ and radiative decay of the QD in non-cavity modes with the rates $\hbar\gamma_\text{rad}^\text{XX}=\SI{1.38}{\mu\eV}$ for XX-X transitions, and $\hbar\gamma_\text{rad}^\text{X}=\SI{1.5}{\mu\eV}$ for X-G transitions, respectively (same rates for H and V polarization). We note here in explicitly, that this are single-transition rates. The XX lifetime is therefore half the inverse of the respective transition decay rate. Beforehand definitions use the common definition of the Lindblad operator for an operator $\hat{A}$,
\begin{equation}
\curlyL_{\hat{A}} \, \rho \ = \ \frac{1}{2} \big( 2 \hat{A} \rho \hat{A}^\dagger - \hat{A}^\dagger \hat{A} \rho - \rho \hat{A}^\dagger \hat{A} \big).
\end{equation}
The sum in \Eqref{RadDec} runs over the annihilation operators of all QD transitions, $\hat{\sigma}=\lbrace\ketbra{\text{G}}{\text{X}_{\text{H,V}}}, \ \ketbra{\text{X}_{\text{H,V}}}{\text{XX}} \rbrace$. Phonon-related contributions in \Eq{\eqref{vNeqAppendix}} are given by
\begin{equation}
    \curlyL_\text{phonons} \, \rho \ = \ \curlyL_\text{deph} \, \rho \ + \ \curlyL_\text{Ph-I} \, \rho. 
\end{equation}
This term contains phonon-induced pure dephasing 
\begin{equation}
    \curlyL_\text{deph} \, \rho \ = \ -\frac{\gamma_\text{deph}}{2} \ \sum_{\hat{\sigma} \neq \hat{\sigma}'} \hat{\sigma} \rho\hat{\sigma}',\label{deph}
\end{equation}
with rate $\hbar \gamma_\text{deph}=\SI{1}{\micro\eV \per \kelvin}\cdot T$ \cite{Laucht_2009,Roy_2011,Praschan_2022}. Here, the sums run over the four electronic states of the QD, $\hat{\sigma},\hat{\sigma}'=\lbrace\ketbra{\text{G}}{\text{G}}, \ \ketbra{\text{X}_{\text{H,V}}}{\text{X}_{\text{H,V}}}, \ \ketbra{\text{XX}}{\text{XX}}\rbrace$. As detailed in \Secref{Sec:QDCavSystem}, we include this term for demonstration purposes. Theoretical results in the main part of the text focus on those for $T=\SI{4.2}{\K}$. The phonon-related contributions further contain phonon-mediated interactions
\begin{equation}\label{LPhI}
\begin{aligned}
    \curlyL_\text{Ph-I} \, \rho \ &= \ -\frac{1}{\hbar^2} \int_0^{\infty}\D \tau \, \sum_{i=g,u}\Big\lbrace G_i(\tau) \times \dotsm \\
    &\dotsm \times \left[ X_i(t),\tilde{X}_i(t,t-\tau)\rho(t-\tau)\right] + \hc \Big\rbrace,
\end{aligned}
\end{equation}
with
\begin{equation}
    X_g \ = \ \chi + \hc \ = \ H_\text{QD-cav} \ + \ H_\text{QD-light},
\end{equation}
and
\begin{equation}
    X_u \ = \ \I (\chi - \hc),
\end{equation}
with use of
\begin{equation}
\begin{aligned}
    \chi \ = \ & \hbar g \sum_{i=\text{H,V}} \big(a_i\ketbra{\text{X}_i}{\text{G}} + a_i \ketbra{\text{XX}}{\text{X}_i}\big) \\
    & - \ \frac{\hbar}{2} \Omega(t) \big( \ketbra{\text{X}_\text{V}}{\text{G}} + \ketbra{\text{XX}}{\text{X}_\text{V}}\big) .
\end{aligned}
\end{equation}
Further, \Eq{\eqref{LPhI}} contains the polaron Green functions
\begin{equation}
    G_g(\tau) \ = \ \AvB^2 \left\lbrace \cosh [\phi(\tau)] -1 \right\rbrace,
\end{equation}
\begin{equation}
    G_u(\tau) \ = \ \AvB^2 \sinh [\phi(\tau)],
\end{equation}
using the phonon correlation function
\begin{equation}
    \phi(\tau) = \int_0^\infty\D\omega \frac{J(\omega)}{\omega^2}\left[\coth\left(\frac{\hbar \omega}{2 k_{\text{B}}T}\right)\cos(\omega \tau) - \I \sin(\omega \tau)\right].
\end{equation}
The propagation in $\tau$ direction of $\tilde{X}_i(t,t-\tau)$ is given by the time evolution operator $U_H(t,t')=\hat{\mathcal{T}}\exp \left[ - \tfrac{\I}{\hbar} \int_t^{t'} \D \mathfrak{t} H(\mathfrak{t}) \right] $: $\tilde{X}_i(t,t-\tau) = U^\dagger_H(t,t-\tau)\tilde{X}_i(t,t)U_H(t,t-\tau)$, with initial condition $\tilde{X}_i(t,t)=X_i(t)$. For visualizing the rate of phonon-mediated cavity feeding in \Figref{Fig2}, we use the analytic approximation of \cite{Roy_2011}. Then, phonon-mediated cavity feeding can be modeled with a Lindblad operator, which depends on the detuning between cavity and respective QD transition, $\Delta=\omega_{\text{Cav}}-\omegaQDT$, and occurs with rate 
\begin{equation}
    \Gamma^{{\rm Cav}, T}_{\rm QD-trans} (\Delta) \ = \ 2\AvB^2 g^2\ \Re\left\lbrace\int \D t \E^{-\I \Delta t} \left(\E^{\phi(t)}-1\right)\right\rbrace. \label{PhononIndQDCavCoupling}
\end{equation}
For numerical efficiency, we evolve \Eq{\eqref{vNeqAppendix}} in the interaction picture by performing transformations like $\tilde{\rho}= U^\dagger \rho U$, where $U=\exp \left[ - \tfrac{\I}{\hbar} (H_\text{QD} + H_\text{cav}) t \right]$. If not mentioned otherwise, we start from an initially excited XX. In the cases where we include TPE, we start from the QD ground state. In any case, we start with empty cavity modes. We solve the polaron master equation explicitly for the system density operator in Fock representation using a 4th order Runge-Kutta method with adaptive step size. The cavity photon numbers are limited to two in each polarization. This is sufficient for the cases investigated due to the exclusion of QD re-excitation by the still high X lifetimes in used cavity configurations. The properties of the cavity photons with polarization $i=\text{H,V}$ are the emission probability \cite{Gustin_2020},
\begin{equation} \label{Eq:CavityEmission}
    \curlyP_{i} \ = \ \kappa \int \D t \langle a_i^\dagger a_i \rangle(t),
\end{equation}
the indistinguishability \cite{Gustin_2018},
\begin{equation}\label{Eq:Indist}
    \curlyI_i \ = \ 1 - \frac{\int \D t \int \D \tau \, 2\GtwoHOMi (t,\tau)}{\int \D t \int \D \tau \, \big( 2 \Gtwopopi (t,\tau) - |\langle a_i(t+\tau)\rangle\langle a_i^\dagger(t)\rangle |^2\big)},
\end{equation}
the purity (definition from \cite{SUPER}, and simplified using \cite{Gustin_2018} similar to $\curlyI_i$),
\begin{equation}
    P_i \ = \ 1 - p_i,
\end{equation}
with
\begin{equation}
    p_i \ = \ \frac{\int \D t \int \D \tau \, \Gtwoi (t,\tau)}{\int \D t \int \D \tau \, \Gtwopopi (t,\tau)}.
\end{equation}
and the concurrence \cite{Wooters_1998,Cygorek_2018,Seidelmann_2021},
\begin{equation}
    \curlyC \ = \ \max \left\lbrace 0,\lambda_4-\lambda_3-\lambda_2-\lambda_1 \right\rbrace.
\end{equation}
Here, $\lambda_i, \, i=1,...,4$ are in rising order the eigenvalues of
\begin{equation}
    R= \sqrt{\sqrt{\twophotmat} \tilde{\rho}^{2\gamma} \sqrt{\twophotmat}},
\end{equation}
where
\begin{equation}
    \tilde{\rho}^{2\gamma} = \sigma_y \otimes (\twophotmat)^\ast \otimes \sigma_y,
\end{equation}
and $^\ast$ denotes the complex conjugate. The elements of the two-photon density matrix $\twophotmat$ calculate as
\begin{equation}\label{2PhotMatAppendix}
    \twophotmat_{ij} = \int \D t \int \D \tau \, \Gtwo_{ij} (t,\tau).
\end{equation}
In \Eq{\eqref{2PhotMatAppendix}} the indices $i,j=1,...4$ represent the two particle linearly polarized basis in standard representation; $\left|1\right\rangle = \left|\text{H,H} \right\rangle$, $\left|2\right\rangle = \left|\text{H,V} \right\rangle$, $\left|3\right\rangle = \left|\text{V,H} \right\rangle$, $\left|4\right\rangle = \left|\text{V,V} \right\rangle$. In that representation, the two particle spin-flip matrix reads
\begin{equation}
    \sigma_y = \left(
 \begin{smallmatrix}
  & & & -1\\
  & & 1 & \\
  & 1 & & \\
  -1 & & & 
 \end{smallmatrix}
\right)
\end{equation}
and the index correspondence between matrix representation and $G^{(2)}$-functions uses $\left|i\right\rangle = \left|i_1,i_2 \right\rangle$ and reads 
\begin{equation}
    \Gtwo_{ij} (t,\tau) \ = \ \Gtwo_{i_1 i_2 j_1 j_2} (t,\tau).
\end{equation}
Calculating entanglement, we consider similar ladder operators as \cite{Bauch_2024}: $b_i=b_i^{\text{XX}}+b_i^{\text{X}}$ with $b_i^{\text{XX}}=a_i+\ketbra{\text{X}_i}{\text{XX}}$ and $b_i^{\text{X}}= \ketbra{\text{G}}{\text{X}_i}$.
The miscellaneous $G^{(i)}$-functions are defined as
\begin{equation}
    \GtwoHOMi (t,\tau) \ = \ \tfrac{1}{2}\big( \Gtwopopi (t,\tau) + \Gtwoi (t,\tau) - |\Gonei (t,\tau)|^2 \big),
\end{equation}
\begin{equation}
    \Gtwopopi (t,\tau) \ = \ \langle a_i^\dagger a_i \rangle (t) \langle a_i^\dagger a_i \rangle (t+\tau),
\end{equation}

\begin{equation}
    \Gtwo_{ijkl} (t,\tau) \ = \ \langle a_i^\dagger (t) a_j^\dagger (t+\tau) a_l(t+\tau) a_k(t)\rangle,
\end{equation}
\begin{equation}
    \Gtwo_{i} (t,\tau) \ = \ \Gtwo_{iiii} (t,\tau),
\end{equation}
\begin{equation}
    \Gonei (t,\tau) \ = \ \langle a_i^\dagger (t) a_i(t+\tau) \rangle. 
\end{equation}
For the evaluation of the miscellaneous $G^{(i)}$-functions we use the quantum regression theorem (QRT) \cite{Carmichael_1999} 
\begin{equation}\label{QRTAppendix}
    \langle \hat{A}(t) \hat{B}(t+\tau) \hat{C}(t)\rangle \ = \ \operatorname{tr} \left\lbrace \Bar{\rho}(\tau)\hat{B} \right\rbrace,
\end{equation}
with
\begin{equation}
    \Bar{\rho}(\tau) \ = \ \left(\hat{C}\rho(t)\hat{A}\right)(\tau)
\end{equation}
for operators $\hat{A}$, $\hat{B}$ and $\hat{C}$. Operators with a time dependence are in the Heisenberg picture, operators without time dependence are their counterparts in the Schr\"{o}dinger picture. Also these calculations are handled in the interaction picture. As given by the QRT, the time evolution of the modified density operator $\Bar{\rho}$ is given by the same equation of motion as for $\rho$, \Eq{\eqref{vNeqAppendix}}. The cavity emission spectrum is calculated using \cite{Mirza_2014}
\begin{equation}
    S_i(\omega) \ = \ \int \D t S_i(t,\omega),
\end{equation}
with the time-dependent spectrum at time $t$
\begin{equation}\label{Eq:Spectrum}
    S_i(t,\omega) \ = \ \kappa \int \D \tau \, \E^{-\I \omega \tau} \Gonei(t,\tau).
\end{equation}
The shorthand notation used in several definitions for double time integrals $\int \D t \int \D \tau$ replaces $\int_0^{t_\text{max}} \D t \int_0^{t_\text{max}-t} \D \tau$. According to this, in the single time integrals $\int \D t$ is the shorthand notation for $\int_0^{t_\text{max}} \D t$, and $\int \D \tau$ is the shorthand notation for $\int_0^{t_\text{max}-t} \D \tau$. The upper limit $t_\text{max}$ is chosen such that the system has fully decayed back to the ground state; $\ket{G,n_\text{H,V}=0}$. 


\section{Derivation of simplified model for resonantly enhanced transition(s) of XX-X cascade}\label{AppX:ExpansionOfCrux}
In this appendix, we first give the connection between the formulas above measuring photon quality in our microscopic theory to the experimentally accessible values. Then, we reiterate the results of previous works \cite{Schoell_2020,Simon_2005}, which present a fundamental limit for the maximal achievable Hong-Ou-Mandel visibility $\curlyV$ depending on the XX-X lifetime ratio. Their theory goes back to \cite{Huang_1993}, where the individual transitions of the XX-X cascade are coupled to flat continua of free photon modes. Wigner-Weisskopf approach leads to a two-photon wave function for the photons generated by the cascade only depending on the lifetimes of the individual transitions. Out of this, \cite{Schoell_2020,Simon_2005} develop their simple limit for $\curlyV$ just depending on the lifetimes. We present an expansion of this theory for the case of Purcell enhancing the individual QD transitions. As a main result, we show that the decay rates (i.e., the inverse of the lifetimes) can simply be replaced by the respective Purcell decay rate $4g^2/\kappa$ for a Purcell enhanced transition \cite{Tomm_2021}. This gives a deeper insight as to why cavity-enhancing the XX-X transition generates high-quality single photons and reflects observable $\kappa$- and $g$-dependencies in this paper. Apart from the high-quality cavity photon emitted from the XX, this also holds true for the X photon. For the different situation of enhancing the X-G transition, our considerations further explain why this particular cavity configuration does not lead to high quality photons.

We start by giving the connection to the visibility $\curlyV$ and $\gtwo$ as measured in the experiment and used in \cite{Schoell_2020,Simon_2005}. The indistinguishability introduced in \Eqref{Eq:Indist} can be rewritten as
\begin{equation}\label{LimitOnI}
    \curlyI \ = \ \frac{1}{2}(1+\curlyV-\gtwo),
\end{equation}
with $\gtwo=\Gtwo/\Gtwopop$ and visibility $\curlyV =\left|\Gone/\Gtwopop\right|^2$. This allows translation of the experimentally measured $\gtwo$ and $\curlyV$ or the limit on the visibility developed in \cite{Schoell_2020,Simon_2005}, 
\begin{equation}\label{LimitOnV}
    \curlyV \ = \ \frac{1}{1+\gamma^{\text{X}}/\gamma^{\text{XX}}},
\end{equation}
into the formalism used in the present paper. For the limit on $\curlyV$ we further assume pure single photons for the individual cascade transitions, $\Gtwo=0$, in $\gtwo=\Gtwo/\Gtwopop$.

We continue developing a theory allowing application of \Eqref{LimitOnV} to the special case of Purcell enhanced transitions. Therefore, we first reiterate the starting point of \cite{Huang_1993}, which is the system Hamiltonian which couples the individual transitions of the cascade to flat continua of photon (bath) modes. Eventually this leads to the two-photon wave function describing the two-photon state emitted by the cascade. Similar to previous works, we focus also on only one polarization channel of the QD. Taking the second polarization channel into account would simply lead to a sum over both polarizations. But all calculations proceed the same. Apart from the free Hamiltonian for QD and bath modes, key ingredient is the coupling Hamiltonian of the QD to flat continua of bath modes
\begin{equation}\label{Eq:QDBath}
    H_\text{QD-bath} \ = \ \hbar \sum_{\omega} \big(\gX \bX\ketbra{\text{X}}{\text{G}} + \gB \bB \ketbra{\text{XX}}{\text{X}}\big) \ + \ \hc .
\end{equation}
Here, $g^i_\omega$, $i={\rm X,XX}$ gives the coupling of the XX-X transition ($i=\rm XX $) or X-G transition ($i=\rm X $) to its respective bath mode with angular frequency $\omega$, described by the annihilation operator $b^i_\omega$. Assuming flat continua for both baths, the index $\omega$ is superficial, but we keep it for easier distinguishability from QD-cavity coupling rates (single mode $\to$ no indexing). For the case of explicitly Purcell enhancing an individual QD transition, this is described again by the dipole interaction Hamiltonian
\begin{equation}\label{Eq:QDCav}
    H_\text{QD-cav} \ = \ \hbar \sum_{\hat{\sigma}} \gsigma \asigma\hat{\sigma}^\dagger \ + \ \hc. 
\end{equation} 
The symbolic QD annihilation operator $\hat{\sigma}$ accounts for the Purcell-enhanced transition and $a^{\hat{\sigma}}$ for the respective cavity mode (resonant with enhanced transition, $\omegasigma$). Coupling is set by $\gsigma$. Therefore, only resonant QD-cavity coupling is considered in this analytics; off-resonant coupling between cavity and the other QD transition is neglected. Furthermore, we neglect direct coupling of the Purcell-enhanced transition $\hat{\sigma}$ to the flat continuum of bath modes as in \Eqref{Eq:QDBath}. This general formalism accounts for various cavity configurations: Enhancing just one of the QD transitions can be achieved by setting ${\hat{\sigma}}=\ketbra{\text{X}}{\text{XX}}$ for enhancing the XX-X transition, or by setting ${\hat{\sigma}}=\ketbra{\text{G}}{\text{X}}$ for enhancing the X-G transition. In both cases, the sum over the symbolic $\hat{\sigma}$ is omitted. Alternatively, one can enhance both transitions selectively with a cavity. Then, the sum over the symbolic $\hat{\sigma}$ accounts for both transitions. Each cavity mode couples to a continuum of free (bath) modes like
\begin{equation} \label{Eq:CavBath}
    H_\text{cav-bath} \ = \ \hbar \sum_{\omega} \Kappasigma \bsigma (\asigma)^\dagger \ + \ \hc .
\end{equation}
Here, $\Kappasigma$ gives the coupling of the cavity mode (enhancing $\hat{\sigma}$-transition) to its respective bath mode with angular frequency $\omega$, described by the annihilation operator $\bsigma$. Again assuming flat continua for the baths, the index $\omega$ is superficial. For consistency, we largely keep the notation of \cite{Huang_1993} in the Hamiltonians in this appendix. Of course, the Lindbladian as introduced in \Eqref{Eq:CavLindbladian} is still obtained with an effective theory for the cavity loss. Therefore, one simply needs to replace the sums with integrals together with $\Kappasigma= -\I \sqrt{\frac{\kappasigma}{2 \pi}}$, where $\kappasigma$ is the loss of the cavity \cite{Gardiner_2004}. But instead of this, we are interested in treating the cavity only implicitly and therefore in an effective Hamiltonian, similar as \Eqref{Eq:QDBath}, where the Purcell enhanced transition $\hat{\sigma}$ couples directly to the cavity bath. The key steps are stated below. One ends up with
\begin{equation}\label{Eq:QDBathEff}
    H_\text{QD-cav-bath} \ = \ \hbar \sum_{\omega} \Gsigma \bsigma \hat{\sigma}^\dagger \ + \ \hc .
\end{equation} 
In this form, the coupling is not a constant anymore but given by 
\begin{equation}
    \Gsigma \ = \ \frac{-\I\gsigma\sqrt{\frac{\kappasigma}{2\pi}}}{\omega-\omegasigma+\I\frac{\kappasigma}{2}},
\end{equation}
whose quadrature gives the Lorentzian shape of the cavity \cite{Michler_2017}. Therefore, the major difference between Purcell-enhanced QD-bath coupling $H_\text{QD-cav-bath}$ in \Eqref{Eq:QDBathEff} and flat QD-bath coupling $H_\text{QD-bath}$ in \Eqref{Eq:QDBath} is the Lorentzian shape of the cavity. But in the Purcell regime ($\kappasigma > \gsigma \gg \gamma_\text{rad}^{\hat{\sigma}}$) where the natural linewidth of the QD transition is much narrower than the cavity, relevant frequencies lie well in the flat region enhanced by the cavity. Therefore, it is allowed to directly apply the flat-continuum theory, even for a Purcell-enhanced transition. Just the coupling needs to be adjusted by $\Gsigma\approx \left. \Gsigma\right|_{\omega=\omega^{\hat{\sigma}}}=-\I \gsigma \sqrt{\frac{2/\pi}{\kappasigma}}$. Therefore, the Purcell-enhanced transition $\hat{\sigma}$ decays with the well known rate $\gamma^{\hat{\sigma}}=4(\gsigma)^2/\kappasigma$ \cite{Tomm_2021}, which consequently replaces its respective not-Purcell-enhanced pendant in \Eqref{LimitOnV}. 

The key steps to derive \Eqref{Eq:QDBathEff} lie in solving the cavity part of the full system Hamiltonian given by \Eqref{Eq:QDCav} and \Eqref{Eq:CavBath} and the free Hamiltonian of QD, cavity and bath. We do this in general for the Purcell enhanced QD-transition $\hat{\sigma}$, as proceedings are independent of the specific QD transition we enhance. For enhancing both transitions with an individual cavity, steps need to be performed for each cavity individually. The cavity dynamics are solved by setting up Heisenberg's equation for the cavity annihilation operator and solving it in frequency space \cite{Gardiner_2004}. To reach a solvable equation, the bath is treated implicitly. Therefore, the free bath Hamiltonian and more importantly the cavity-bath-interaction Hamiltonian are replaced by the Lindbladian $\kappasigma\curlyLTilde_{\asigma}$, with $\curlyLTilde_{\hat{A}} \, \hat{X} = \frac{1}{2} \big( 2 \hat{A}^\dagger \hat{X} \hat{A} - \hat{A}^\dagger \hat{A} \hat{X} - \hat{X} \hat{A}^\dagger \hat{A} \big)$. Then, the Heisenberg equation reads
\begin{equation}\label{eq:Heisenberg}
    {\frac{\D}{\D t}}\asigma \ = \ \tfrac{\I}{\hbar} \, \big[ H , \rho \big] \ + \ \kappasigma\curlyLTilde_{\asigma} \, \asigma.
\end{equation}
Evaluating all contributions leads to
\begin{equation}
    {\frac{\D}{\D t}}\asigma \ = \ - \I \omegasigma \asigma - \I \gsigma \hat{\sigma} -\frac{\kappasigma}{2}\asigma.
\end{equation}
In frequency space this reads
\begin{equation}
    -\I\omega\asigma(\omega) \ = \ - \I \omegasigma \asigma(\omega) - \I\gsigma \hat{\sigma}(\omega) -\frac{\kappasigma}{2}\asigma(\omega).
\end{equation}
The algebraic solution for the cavity annihilation operator reads
\begin{equation}\label{eq:SolutionCAO}
    \asigma(\omega) \ = \ \frac{\gsigma}{\omega-\omegasigma+\I\frac{\kappasigma}{2}} \, \hat{\sigma}(\omega).
\end{equation}
To insert this in \Eqref{Eq:CavBath} and to derive \Eqref{Eq:QDBathEff}, one needs to perform the inverse Fourier transform using the requirement that the QD linewidth is narrow compared to the cavity linewidth, $\gamma_\text{rad}^{\hat{\sigma}} \ll \kappasigma$. The free Hamiltonian of the cavity and QD-cavity-interaction Hamiltonian can be neglected in the Purcell regime, as photons immediately leave the cavity. Nonetheless, these terms would only generate a phase which proved to be incidental in the context of lifetime ratio limiting the visibility.


\section{Derivation of simplified model for general case of naturally decaying XX-X cascade in a tunable cavity}\label{AppX:ConnectionToExp} 
Here we derive a generalized version of the theory presented in the preceding section. To this end we simultaneously consider the coupling of each QD transition to both a cavity (not necessarily on resonance) and a flat continuum. This approach makes it possible to study situations with smaller QD-cavity coupling, for which the corresponding Purcell rates are of a magnitude comparable to the radiative decay rate. Whereas the theory presented earlier approximates the radiative decay (in non-cavity modes) as negligible for a cavity-enhanced transition, the extended theory introduced here removes this limitation and thus provides more accurate results, including the cases where radiative decay has only vanishing impact. This increased generality comes at the cost of requiring additional steps and assumptions. For this reason, both theoretical descriptions are included in the present paper. 

As a very foundational result of the present paper, the key achievement of this most general derivation is that it enables a detailed description of the experimental situation for the present paper and also that of Ref.~\cite{Baltisberger_2025}. This involves a tunable cavity coupled to the radiatively decaying QD transitions. Our experiments \cite{Baltisberger_2025} turn out to lie just within the validity range of the approximations assumed for the analytical derivation below, explaining the excellent agreement of the measurements with the simple analytical result based on the lifetime ratio $\tau_{\mathrm{XX}}/\tau_{\mathrm{X}}$ as plotted in \Figref{Fig1}\nolinebreak (c).

Let us consider the simultaneous coupling of a transition, represented by the operator $\hat{\sigma}$, to a cavity of angular frequency $\omegaCav$ and to a flat continuum. The cavity is again coupled to only one QD transition. All previous definitions remain unchanged, except that the cavity is not necessarily resonant with the transition (still at $\omegasigma$), which requires the extension of our notation. Similarly, we adapt the notation for the cavity loss $\kappacav$ and for the cavity CAOs, $\acav^{(\dagger)}$. The cavity is coupled to a continuum of free (bath) modes as described in \Eqref{Eq:CavBath}. All Hamiltonians remain the same, with the only difference being that the Hamiltonians associated with both processes -- the QD-continuum coupling (see \Eqref{Eq:QDBath}) and the QD-cavity-continuum coupling (see \Eqsref{Eq:QDCav} and \ref{Eq:CavBath}) -- are now considered simultaneously in an additive fashion. The elimination of the explicit treatment of the cavity can be carried out using the same procedure as before: the steps presented from \Eqref{eq:Heisenberg} to \Eqref{eq:SolutionCAO} can be repeated for an arbitrary cavity frequency. One ends up with
\begin{equation}\label{Eq:QDBathEffAdv}
    \tilde{H}_\text{QD-cav-bath} \ = \ \hbar \sum_{\omega} \GsigmaTilde \bsigma \hat{\sigma}^\dagger \ + \ \hc ,
\end{equation} 
where the newly introduced couplings $\GsigmaTilde = \gsigmaConti + \Gsigma$ combine the direct QD-continuum coupling $\gsigmaConti$ and the cavity-enhanced QD-continuum coupling $\Gsigma$. Definitions remain the same, but in $\Gsigma$ the cavity frequency needs again to be replaced by its general version $\omegaCav$, as resonance with the QD transition is not necessarily provided. In \Eqref{Eq:QDBathEffAdv}, we have united the initially separately treated bath operators of the QD transition and the cavity to bath operators $\bsigma$ (same bath of free photon modes). All steps performed until now can be carried out for both QD transitions, if both have QD-continuum and QD-cavity-continuum coupling simultaneously. Repeating the Wigner-Weisskopf approach presented in \cite{Huang_1993} shows that \Eqref{LimitOnV} remains valid in the Purcell regime, but with decay rates containing all possible decay paths of transition $\hat{\sigma}$, 
\begin{equation}\label{Eq:FullDecayRates}
\gamma^{\hat{\sigma}}=\gamma_\text{Purcell}^{\hat{\sigma},\omegaCav}+\gamma_\text{rad}^{\hat{\sigma}}.
\end{equation}
Importantly, the Purcell rate now contains the Lorentzian shape, 
\begin{equation}
\gamma_\text{Purcell}^{\hat{\sigma},\omegaCav} \ = \ \frac{(\gsigma)^2\kappacav}{(\omegasigma-\omegaCav)^2+(\frac{\kappacav}{2})^2} ,
\end{equation}
as QD transition and cavity are not necessarily resonant. Also, outside the cavity resonance the QD transition only experiences the local value of the Lorentzian, provided that the QD transition linewidth is narrow compared to the cavity linewidth, $\gamma_\text{rad}^{\hat{\sigma}} \ll \kappacav$. Therefore, the effective QD-cavity-continuum coupling can again be approximated as flat for the relevant frequencies. However, the coupling is reduced by the Lorentzian in comparison to the resonant case.

This result now allows us to fully capture the experimental situation in Ref.~\cite{Baltisberger_2025}. The crucial difference is that both transitions of the XX-X cascade couple to the same cavity. This translates to the evaluated Heisenberg equation for the cavity CAOs
\begin{equation}\label{eq:HeisenbergAdv}
    {\frac{\D}{\D t}}\acav \ = \ - \I \omegaCav \acav - \I \sum_{\hat{\sigma}} \gsigma \hat{\sigma} - \frac{\kappacav}{2}\acav,
\end{equation}
where the sum runs over the annihilation operators of both QD transitions, $\hat{\sigma}=\lbrace\ketbra{\text{X}}{\text{XX}},\ketbra{\text{G}}{\text{X}}\rbrace$. The algebraic solution in frequency space now reads
\begin{equation}\label{eq:SolutionCAOSingleCav}
    \acav(\omega) \ = \ \sum_{\hat{\sigma}}\frac{\gsigma}{\omega-\omegaCav+\I\frac{\kappacav}{2}} \, \hat{\sigma}(\omega).
\end{equation}
Performing the same steps as those leading to \Eqref{Eq:QDBathEffAdv} may become crucial, since both QD transitions are coupled to the same cavity. Consequently, both QD transitions (as well as the cavity itself) couple to the same bath of free photon modes. This situation appears to eliminate the initial distinction between photons originating from individual QD transitions. However, as long as the QD transitions remain spectrally separated, and as long as the QD-cavity-continuum coupling $\Gsigma$ is Lorentzian-shaped, it is justified to again treat the photons associated with each transition independently. Therefore, we can write
\begin{equation}\label{Eq:QDBathEffAdvSingleCav} 
    \tilde{H}_\text{QD-cav-bath} \ = \ \hbar \sum_{\hat{\sigma}}\sum_{\omega} \GsigmaTilde \bsigma \hat{\sigma}^\dagger \ + \ \hc ,
\end{equation}
and the previously derived results remain applicable, even when both QD transitions couple to the same cavity. This configuration corresponds precisely to the experimental situation involving a single tunable cavity coupled to both QD transitions of the XX-X cascade. We have therefore demonstrated that \Eqref{LimitOnV} applies to the collective decay rates encompassing all available decay channels of a given transition, or, equivalently, to the (measured) lifetimes of the XX and X states.

Finally, we note that the discussion here focuses solely on the essential part of the experiment, which can be reduced to a three‑level cascade (a single X state) and a single cavity mode (one polarization). The second X state and the corresponding cavity mode can be neglected, since the cavity is detuned in frequency. Hence, their influence is limited to the emission rather than the visibilities. The presence of the other X only needs to be taken into account in the lifetime of the XX. For demonstration of this analytic model we refer to \Appref{AppX:DiffTemp}, \Figref{FigA1} and \Figref{FigA3}. Good agreement with the numerical results is observed in the validity range of the analytical model. Further discussion is given there.


\section{Further enhancing the single-photon character by filtering}\label{AppX:SpectralFiltering}
In this appendix, we investigate the influence of spectral filtering on the quality of the cavity output, using the sensor method \cite{DelValle_2012,Bermudez_2025}. We begin by introducing the theory and methods used, and then present the results.

For filtering the output of the H-cavity mode at angular frequency $\omegaS$ with linewidth $\muS$, we couple it to a sensor positioned at respective frequency with CAOs $S^{(\dagger)}$,
\begin{equation}
    H_\text{cav-sensor} \ = \ \hbar \epsiS a_{\text{H}}S^\dagger \ + \ \hc .
\end{equation}
The coupling $\epsiS$ needs to be chosen small enough ($\epsiS \ll \sqrt{\muS \kappa}$), to avoid any back action on the cavity. In particular, we use $\epsiS= \tfrac{\kappa}{100}$. The sensor linewidth corresponds to the linewidth of the filter $\muS$ and will be varied in our analysis. To implement the filter, we add $H_\text{cav-sensor}$ and the free evolution of the sensor
\begin{equation}
    H_\text{sensor} \ = \ \hbar \omegaS S^\dagger S
\end{equation}
to the system Hamiltonian. Further, we need to add the Lindbladian
\begin{equation}\label{Eq:SensorLindbladian}
    \curlyL_\text{sensor} \, \rho \ = \ \muS \ \curlyL_{S} \, \rho.
\end{equation}
To minimize computational complexity in these investigations, we handle the direct coupling of the QD to the V-cavity mode (detuned from all QD transitions) by representing it solely through effective Lindblad terms
\begin{equation}\label{Eq:VCavLindbladian}
    \curlyL_\text{V-Cavity} \, \rho \ = \ \gamma^{\text{Purcell}}_{\text{XX}\to\text{X}_\text{V}} \ \curlyL_{\ketbra{\text{X}_\text{V}}{\text{XX}}} \, \rho \ + \ \gamma^{\text{Purcell}}_{\text{X}_\text{V}\to\text{G}} \ \curlyL_{\ketbra{\text{G}}{\text{X}_\text{V}}} \, \rho.
\end{equation}
Here, we use the Purcell rate
\begin{equation}\label{Eq:Purcell}
    \gamma^{\text{Purcell}}_{\text{QD-trans}} \ = \ \frac{g^2\kappa}{(\omegaQDT-\omega_{\text{V}})^2+(\frac{\kappa}{2})^2} 
\end{equation}
with the respective QD-transition frequency $\omegaQDT=\omega_{\text{XX}\to\text{X}_\text{V}}=(E_{\text{XX}}-E_{\text{X}_\text{V}})/\hbar$ for $\gamma^{\text{Purcell}}_{\text{XX}\to\text{X}_\text{V}}$, and $\omegaQDT=\omega_{\text{X}_\text{V}\to\text{G}}=(E_{\text{X}_\text{V}}-E_{\text{G}})/\hbar$ for $\gamma^{\text{Purcell}}_{\text{X}_\text{V}\to\text{G}}$. For phonon-mediated cavity feeding of the V mode we proceed the same and add the Lindblad terms
\begin{equation}\label{Eq:VCavPhononLindbladian}
    \curlyL^{\text{Phonons}}_\text{V-Cavity} \, \rho \ = \ \Gamma^{{\rm Cav}, T}_{\text{XX}\to\text{X}_\text{V}} \ \curlyL_{\ketbra{\text{X}_\text{V}}{\text{XX}}} \, \rho \ + \ \Gamma^{{\rm Cav}, T}_{\text{X}_\text{V}\to\text{G}} \ \curlyL_{\ketbra{\text{G}}{\text{X}_\text{V}}} \, \rho,
\end{equation}
using the analytical rate $\Gamma^{{\rm Cav}, T}_{\rm QD-trans} (\Delta)$ as introduced in \Eqref{PhononIndQDCavCoupling} for $\omega_{\text{Cav}}=\omegaV$ and $\omegaQDT= \omega_{\text{XX}\to\text{X}_\text{V}}$ or $\omegaQDT=\omega_{\text{X}_\text{V}\to\text{G}}$, respectively. We note here explicitly, that our effective treatment of phonon-mediated cavity feeding is different from \cite{Roy_2011} by neglecting the V-cavity mode CAOs in the Lindblad terms. This is valid for all situations considered in this appendix, as we are not explicitly interested in the V-cavity photons, and their photon numbers are much lower than unity. Therefore, the common factors $\sqrt{n}$ and $\sqrt{n+1}$ introduced by the CAOs are simply one. In general, our effective treatment of the V-cavity mode has no influence on the single-photon character of photons in the H-cavity mode, as in the studied parameter regimes the presence of the V mode generally only tangents the emission. Insignificant discrepancies in the emissions (almost always significantly below $\SI{1}{\percent}$) can be explained by the different models used to describe the V mode. 

The photon quality quantifiers of the filtered signal are now calculated using the formulas introduced in \Appref{AppX:Theory}, but with the sensor CAOs $S^{(\dagger)}$ instead of the CAOs for the H-cavity mode. To arrive at the filtered emission and cavity emission spectrum, apart from the CAOs also the loss rates need to be replaced in \Eqsref{Eq:CavityEmission} and \eqref{Eq:Spectrum} by $\kappa \left({\muS}/{2\epsiS}\right)^2$. Again, we perform all calculations in the interaction picture with methods introduced in \Appref{AppX:Theory}.

\begin{figure}
\centering
\includegraphics[width=0.8\columnwidth]{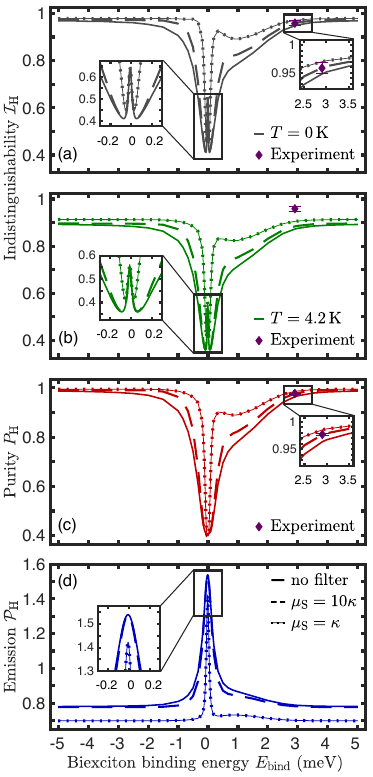} 
\caption{{Partial restoration of single-photon character for different biexciton binding energies with spectral filtering.} Shown are for (filtered) H-mode cavity photons (a) photon indistinguishability $\IH$ for $T=\SI{0}{\K}$ (dark gray), and (b) photon indistinguishability $\IH$ (green), (c) single-photon purity $\PH$ (red), (d) emission $\EH$ (blue) for $T=\SI{4.2}{\K}$ for different filter widths for initially excited biexciton and empty cavity. Results without filtering (obtained with cavity CAOs) are shown as solid lines, with filtering widths $\muS=10\kappa$ (dashed lines) and $\muS=\kappa$ (pearl chains). Results of experimental realization \cite{Baltisberger_2025} are included as purple diamonds with error bars in (a-c). Detuning between X-G transition and cavity changes with biexciton binding energy as $\hbar\DeltaXHG = -\Ebind-\Efsp \approx -\Ebind$. Results without filtering correspond to the results presented in \Figref{Fig2}.}\label{FigA4}
\end{figure}

In the following, we present the results for different filter widths $\muS$. As the XX binding energy has proven its crucial role for our emission scheme (see \Secref{Sec:GeneralPotential}, \Figref{Fig2}), we analyze to what extend this can be minimized by filtering at the cavity frequency, $\omegaS=\omegaH$. Therefore, the negative influence of the additional X photons emitted into the cavity is expected to be reduced. More specifically, this way the X photons emitted off-resonantly in the cavity can be erased from the cavity output. Those photons correspond to the side peak of the cavity emission spectra, as visible in \Figref{Fig1}\nolinebreak (b). Not erased by this method will be the X photons emitted resonantly into the cavity, i.e., by phonon-mediated cavity feeding. Therefore, we only expect a reduction, but not total disappearance of the negative influence of the X photons on the photon output by using spectral filtering. As for larger absolute XX binding energies the negative influence of the X photons has proven to disappear, photon qualities observable there can be used as benchmark for the improvement through spectral filtering. The sensor method corresponds to filtering with a Lorentzian mask. Therefore, also the cavity broadened XX emission and therefore the major part of the cavity output might be reduced by narrower filters. Ideally spectral filtering is only applied in moderation such that it leaves the amount of XX-cavity photons emitted intact.

\begin{figure}
\centering
\includegraphics[width=0.95\columnwidth]{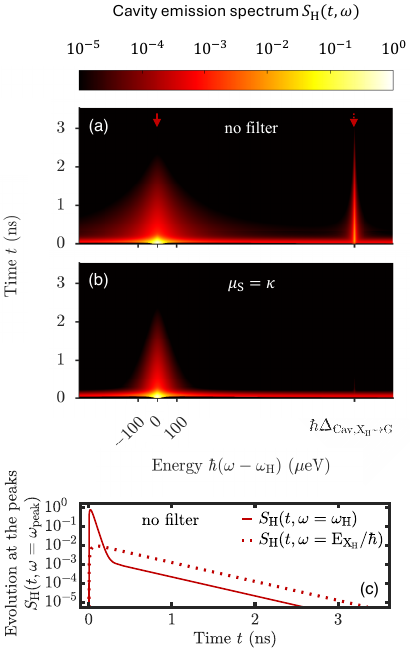} 
\caption{{Influence of spectral filtering on time-dependent cavity emission spectra.} Shown are the H-mode time-dependent cavity emission spectra $S_{\text{H}}(t,\omegaH)$ without filtering (a) and with narrow filtering (b) ($\muS=\kappa$) for $\Ebind=\SI{1}{\meV}$ for better illustration and $T=\SI{4.2}{\K}$. The time evolutions of the peaks of the unfiltered spectrum as shown in (a) (marked with arrows) are further shown as line plots in (c). The peak at cavity frequency $\omegaH$ is shown as solid line, the peak at H-exciton frequency $E_{\text{X}_\text{H}}/\hbar$ is shown as dotted line. Results are for initially excited biexciton and empty cavity. Results without filtering are obtained with cavity CAOs.}\label{FigA5} 
\end{figure}

\FIGref{FigA4} shows our results for the photon qualities without any filtering (solid lines), for a filter broader than the cavity width ($\muS=10\kappa$, dashed lines), and for a filter fitting the cavity width ($\muS=1\kappa$, pearl chains) in dependency of the XX binding energy. Colors as before; indistinguishability $\IH$ is shown in green, purity $\PH$ in red, and emission $\EH$ in blue. Results presented in colors are for $\SI{4.2}{\K}$, indistinguishabilities presented in dark gray are for $\SI{0}{\K}$. For conciseness we do not show the results of the phonon free case for $\SI{0}{\K}$. Just as without spectral filtering (\Figref{Fig2}) this case shows similar behavior for positive and negative XX binding energies. The phonon-free case behaves for both signs like the case of $\SI{0}{\K}$ (including phonons) at negative XX binding energies, where almost no phonon influence is visible. Also, we do not show purity and emission for the case of $\SI{0}{\K}$, as they also only slightly deviate from $\SI{4.2}{\K}$. Overall, we observe that spectral filtering slightly reduces the emission and increases the single-photon character. Both effects are generally more pronounced with the narrower filter. 

As expected, also the time resolved cavity emission spectra in \Figref{FigA5} reflect our observation. For reasons of clarity and visualization, results are shown for $\Ebind=\SI{1}{\meV}$ and $T=\SI{4.2}{\K}$ with the narrow filter. The side peak caused by the X is reduced by spectral filtering. Note that the X photons emitted resonantly into the cavity, i.e., by phonon-assisted cavity feeding, are not removed by spectral filtering. 

Focusing on the time evolutions of the unfiltered peaks as shown in \Figref{FigA5} (c), it is expected that temporal filtering may further boost the photon quality. While the XX (and the associated intense peak in the spectrum) is emitted on a short timescale of a few tens of picoseconds due to Purcell enhancement, the X and associated peaks (the tail of the intense peak at cavity resonance and the side peak) decay on much longer timescales (a few hundred picoseconds). Therefore, XX-cavity photons can be temporally separated from the major part of the X photons. In any case we would like to reiterate that spectral filtering will not alleviate the strict physical limit for the indistinguishability set by the lifetime ratio, nor eliminate phonon-induced pure dephasing.

As also discussed in the main part, for the relevant performance quantifiers of our single-photon source we obtain for $\Ebind=\SI{5}{\meV}$ at $T=\SI{0}{\K}$ with moderate spectral filtering ($\muS=\kappa$) values of $\gtwo=0.0041$ and $\curlyV=0.9607$. Again, the case of $T=\SI{0}{\K}$ is chosen here as it best approximates our present experiment (see \Figref{Fig2} and \Figref{FigA4}) with very low noise and transitions close to the transform limit \cite{Baltisberger_2025}.


%

\end{document}